\documentclass[pdflatex,sn-mathphys-num]{iopjournal}
\usepackage{graphicx}%
\usepackage{svg}
\svgsetup{inkscapelatex=true}

\usepackage[utf8]{inputenc}
\usepackage{newunicodechar}
\newunicodechar{ }{\,}
\usepackage{tcolorbox}
\usepackage{caption}
\usepackage{subcaption}

\usepackage{multirow}%
\usepackage{amssymb,amsfonts}
\usepackage{amsthm}%
\usepackage{mathrsfs}%
\usepackage[title]{appendix}%
\usepackage{xcolor}%
\usepackage{textcomp}%
\usepackage{manyfoot}%
\usepackage{booktabs}%
\usepackage{float}
\usepackage{harvard}
\citationmode{abbr}
\usepackage[export]{adjustbox}
\usepackage{gensymb}

\usepackage{parskip}
\usepackage{algorithm}%
\usepackage{algorithmicx}%
\usepackage{algpseudocode}%
\usepackage{listings}%
\usepackage{siunitx}
\begin{document}

\articletype{Paper}

\title{Magnet-Free Proton Therapy with 4D Pencil Beam Delivery Optimisation}

\author{Nair N von Mühlenen$^1$$^*$, Florentin Bieder $^1$, Ye Zhang $^2$ and Philippe C Cattin$^1$}

\affil{$^1$ Department of Biomedical Engineering, University Basel, 4001 Basel, Switzerland}

\affil{$^2$ Center for Proton Therapy, Paul Scherrer Institut, 5232 Villigen, Switzerland}

\affil{$^*$ Author to whom any correspondence should be addressed.}

\email{nair.vonmuehlenen@unibas.ch}


\date{August 2025}

\begin{abstract}
\textit{Objective}. Motion management is a critical challenge in proton therapy for mobile tumours. This study aims to develop and evaluate a novel four-dimensional (4D) pencil beam delivery strategy that incorporates respiratory motion into a dynamic treatment plan to improve dose conformity and treatment efficiency.
\textit{Approach}. To assess this 4D pencil beam delivery strategy, a mobile phantom was used. The generated 4D treatment plans were assessed with various scanner configurations, including gantry-free and magnet-free scanner heads. For each setup, the treatment time, dose conformity, and robustness against irregular breathing patterns were quantified. The influence of scanner head design and patient-specific motion irregularities on overall plan quality was evaluated.
\textit{Main Results}. The 4D planning tool generated treatment plans that achieved clinically acceptable dose distributions across all configurations. In magnet-free configurations, static beam operation substantially increased treatment time and reduced dose conformity. In contrast, configurations using a single scanner magnet, without a gantry, maintained acceptable conformity within practical treatment times.
\textit{Significance}. The proposed 4D delivery strategy demonstrates feasibility for treating mobile targets with simplified, gantry-free and magnet-free scanner designs. Further improvements could be achieved by synchronising the patient's breathing with 4D delivery, which may enhance dose accuracy during irregular or interrupted breathing. By reducing system complexity while preserving dosimetric performance, this approach offers a pathway toward more accessible and cost-effective proton beam therapy for motion-affected tumours.

\end{abstract}

\section{Introduction}
Cancer is one of the leading causes of death worldwide. Alone in 2022, the estimate of cancer deaths was around 9.7 million, accompanied by around 20 million new cancer cases.\cite{bray2024global}. For those diagnosed with cancer, there are many possible treatment options, such as chemotherapy, conventional radiotherapy, or proton radiotherapy (PT). PT has emerged as a powerful treatment tool \cite{smith2006proton}. The unique physical properties of protons, particularly the Bragg Peak, enable highly localised radiation deposition \cite{schulz2007particle}. This precision is especially beneficial for tumours near sensitive organs, allowing you to avoid irradiating organs at risk (OAR) and, in general, reducing the overall integral dose. Although this property is favourable for shielding both vulnerable and healthy tissues from excessive radiation, it also makes PT more susceptible to motion. Currently, no devices verify during treatment if the radiation is delivered to the intended position. This can lead to underdosed areas of the tumour and an overdose within healthy tissue instead. Strategies like gating or motion-encompassing margins are currently being used to mitigate some of the motion uncertainty.

Nevertheless, PT is associated with fewer general side effects than other cancer treatment options, making it a favourable option, especially in paediatric oncology. 
The Centre for Proton Therapy (CPT) at the Paul Scherrer Institute (PSI) states that approximately 36\(\%\) of patients are under the age of 18 \cite{noauthor_patients_nodate}. 
It is essential to note that potential side effects are highly dependent on the treatment site and the patient's individual circumstances, and as of 2025, still lack large-scale investigative studies \cite{salem2024proton}. The lack of studies is due to the inaccessibility of PT in many regions of the world, particularly in economically weaker countries. This lies mostly in the high cost of building and maintaining a PT centre. According to the Particle Therapy Co-Operative Group (PTCOG), 135 facilities worldwide are currently in use \cite{noauthor_ptcog_nodate}. Most of these facilities are located in the USA (48) and Japan (26), but there is no PT Centre on the entire African continent.  

In this work, we aim to address these two problems: PT's high motion sensitivity and the inaccessibility due to high cost. We present a new motion management tool, 4D pencil beam delivery optimisation, which enables treatment planning for moving targets. 
In addition, we investigate how we could leverage this tool to work with gantry-less and magnet-free proton delivery systems, including \emph{self-scanning} \cite{Moeri2016,Moeri2017}, i.e., leveraging the respiratory motion for passive scanning of the target. 
We test the 4D treatment plans using a 3D phantom representing a liver tumour under motion, and assess the treatment outcome under regular and irregular breathing conditions. 

\subsection{Motion Mitigation}
The general idea of motion management is to mitigate or minimise the effect of intra- and inter-fractional motion \cite{smith2006proton,schulz2007particle}. Intra-fractional motion can be caused by respiration and, to a lesser degree, by peristaltic and cardiac motion. Inter-fractional motion refers to changes due to tumour shrinkage, variations in the volumes of hollow organs \cite{wang2016assessment}, or weight loss. Inter-fractional motion usually happens slowly over days, allowing it to be accounted for if correctly identified. Intra-fractional motion occurs during treatment in timespans of seconds to minutes \cite{SIEBENTHAL2007}, posing a greater challenge. Motion management or mitigation strategies can be divided into techniques that prevent or reduce anatomical motion and those that adapt treatment planning or delivery to motion \cite{pakela2022management}.

\subsubsection*{Dose Delivery}
Respiratory gating and repainting are mitigation options that are used during treatment delivery. Respiratory gating is a dynamic delivery method, in which the beam is activated only during a specific phase of the breathing cycle, e.g., at the end of exhalation \cite{ebner2017respiration,gelover2019clinical}. Respiratory gating ensures a more consistent dose delivery, minimising the risk of under- or overshooting the target. Repainting involves rescanning the same position multiple times, increasing the statistical likelihood of accurately treating the planned position. It mitigates additional dose uncertainty caused by the interplay effect between the patient's breathing pattern and the scanning sequence \cite{zenklusen2010study}.

\subsubsection*{Immobilization}
A standard practice during treatment is the immobilisation of the patient using patient-specific cushioning, head- and mouthpieces \cite{wroe2015clinical}. In the thorax and upper region of the abdomen, the most prevalent motion is breathing. A standard technique for minimising respiration-based motion is breath-hold, where the patient holds their breath for a specific period of time \cite{dueck2016robustness}, or breathing is actively controlled via a ventilator \cite{wong1999use}. Another method to reduce intra-fractional motion is the use of an abdominal compression belt \cite{lin2017evaluation}. It is inexpensive and easily implemented. However, the amount of applicable pressure depends on the patient's tolerance, resulting in variable reductions in intra-fractional motion.

\subsubsection*{Planning}
With appropriate treatment planning, it is possible to account for some of the
intra-fractional motion during delivery. Computed Tomography (CT) images are the gold standard for planning PT \cite{balter1996uncertainties}. It is possible to capture the range of breathing motion in CT scans acquired at different phases of the respiratory cycle. These 4D CT scans, however, are susceptible to breathing irregularities during acquisition or day-to-day variability in breathing patterns \cite{SIEBENTHAL2007,sindoni2016usefulness,krieger2020impact}. This can lead to an inaccurate representation of the patient's respiratory motion. To account for these irregularities, various methods have been designed. A widely adopted strategy is to add motion-encompassing margin around the target volume, also known as the \emph{internal target volume}, during the planning phase \cite{van2000probability,czerska2021clinical}. It increases the probability of treating the entire target, with the cost of increasing the dosage within healthy tissue. Generally, motion-encompassing margins are used in conjunction with other mitigation or management strategies to ensure that most motion uncertainties are considered. 

Appropriate beam angle selection is another strategy used during the planning phase to minimise the effect of breathing motion. The proton range is sensitive to changes in density. Variation in tissue density along the beam path due to breathing can affect the beam's trajectory. At certain angles, the organs' motion is less pronounced, minimising tissue density gradients. Thus, choosing such beam angles reduces uncertainties in dose deposition \cite{chang2017consensus}. More modern approaches model the patient's breathing pattern using machine learning, achieving great success, utilising ultrasound or beam-eye-viewer X-ray images to predict organ movement during breathing \cite{Preiswerk2014,giger2018respiratory,giger2019inter,xia2025novel}. 

\subsection{Gantry-less Proton Therapy}\label{section:GantryLess}
Solutions for delivery systems without a gantry have been investigated for a considerable time. The \emph{Planar Iso-centric System} is a three-headed beam line system that replaces the gantry while retaining some rotational \emph{DOs}. It is also possible to compensate for the lack of rotational \emph{DOF} with an adjustable patient positioning system \cite{feldman2024commissioning}. For example, patients are positioned on a turning platform or chair with the beam set at a fixed angle \cite{kats2017gantry}. Having the patient rotate instead of the beam. A completely different approach is to extend a gantry-less proton setup with photon therapy \cite{fabiano2020combined}. The lack of \emph{DOFs} in the PT setup compromises the treatment quality. By supplementing with photon beams, the accuracy and quality of therapy can be improved.

\section{Method}
For the 4D pencil beam delivery optimisation, we utilised three phantoms with varying target volumes. For each target volume, we generated a 4D delivery plan under regular breathing conditions for spot-scanning PT. The plans contain the coordinates of each spot, the delivery time and optimised beam weights. These plans were adapted and then optimised for different beam delivery systems, including gantry-free and magnet-free systems. We simulated the plans under both regular and irregular breathing conditions and evaluated them with metrics generally used in PT.

\subsection{Treatment Simulation}\label{section:TreatmentSimulation}
\begin{figure} [H]
    \centering
    \begin{subfigure}{0.4\textwidth}
        \centering
         \includegraphics[width=0.8\textwidth]{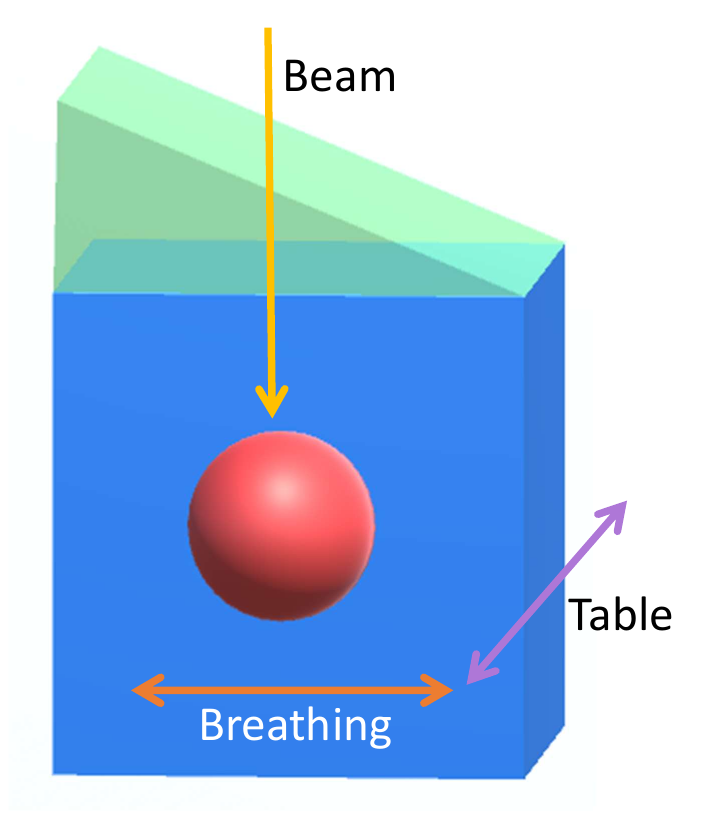}
        \caption{Schematic of the phantom.}\label{fig:phantom} 
    \end{subfigure}%
    \vskip\baselineskip 
    \begin{subfigure}{\textwidth}
         \includegraphics[width=0.8\textwidth]{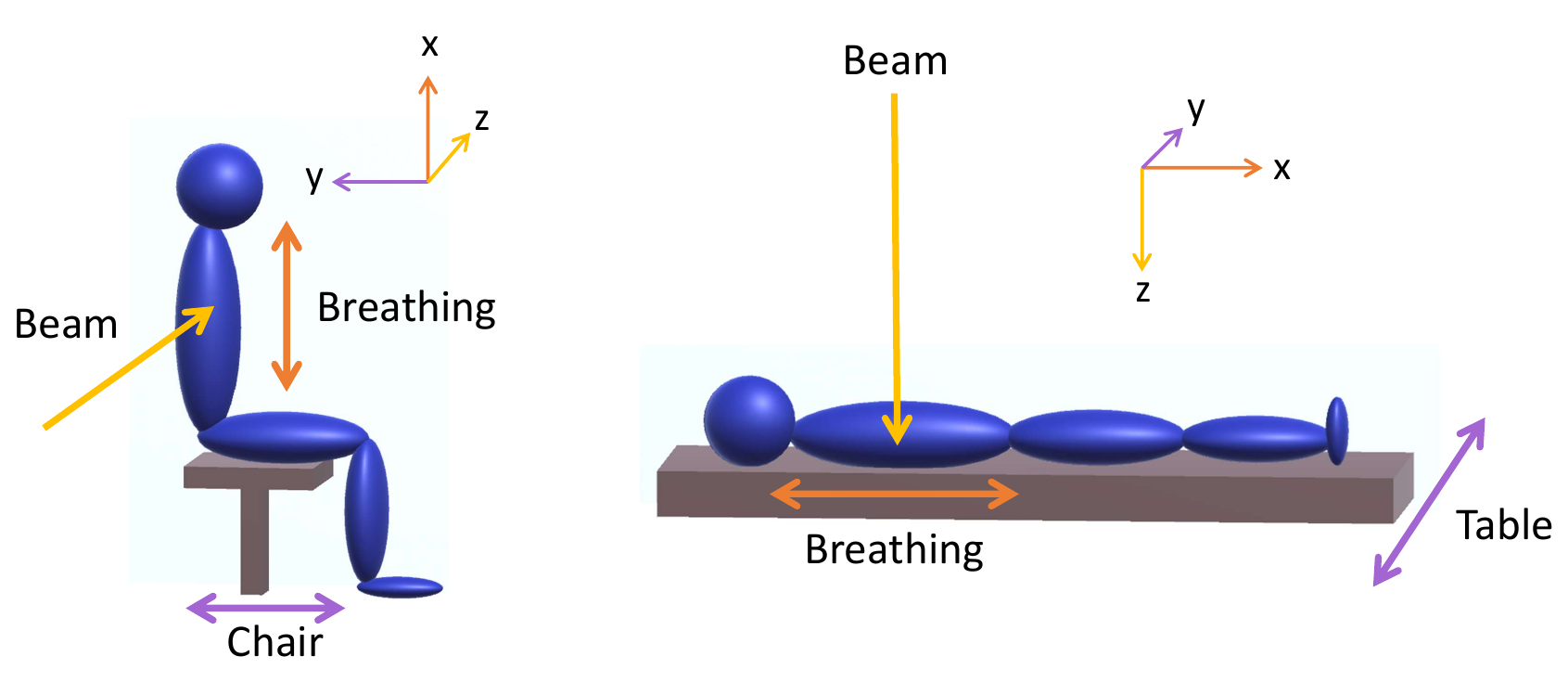}
        \caption{Patient coordinate systems}
        \label{fig:patient}
    \end{subfigure}
    \caption{Schematic of the phantom and the corresponding patient coordinate system. (a) The phantom consists of a target volume representing a liver tumour (red sphere), the liver (blue rectangle) and a rib bone (green wedge). The surrounding tissue can represent water or fatty tissue. (b) Patient coordinate system for sitting and lying patients with respect to the beam delivery system. The beam always lies along the $z$-axis.} \label{fig:ppp}
\end{figure}

We chose a phantom setup that incorporated different types of tissues with varying thicknesses. Figure~\ref{fig:ppp} shows a schematic of a phantom (Figure~\ref{fig:phantom}) and the corresponding patient coordinate system (Figure~\ref{fig:patient}). The red sphere represents a liver tumour, the blue rectangle represents the liver, and the transparent green wedge represents a rib bone. The surrounding tissue can be water or fatty tissue. The yellow arrow indicates the direction of the beam, the orange arrow the breathing motion, and the purple arrow a possible table or chair motion. It is essential to note that the $z$-axis always lies along the beam direction. The phantom displays each tissue with its respective Hounsfield units (HU) \cite{schaffner1998precision}. The phantom settings allow you to change the size, shape, and HU values of each component. For our experiments, we chose three target volumes with diameters of \SI{30}{mm}, \SI{40}{mm} and \SI{50}{mm}. Breathing is approximated with a $\cos^{4}$ function \cite{fukumitsu2012reproducibility,pleil2021physics} and affects only the blue rectangle and the red target sphere. All the parameters used for the breathing modelling can be found in Table~\ref{tab:PhantomVariables} and were chosen according to the average breathing of an adult human. The values are adaptable and can thus simulate different breathing patterns. To mimic irregular breathing, both the amplitude and frequency are modulated over time. 

\begin{table*}[h]
\centering
\caption{Phantom-specific breathing time and depth.}\label{tab:PhantomVariables}
\begin{tabular}{lll}
        \toprule
        \textbf{Notations} & \textbf{Definition} & \textbf{Variable}\\
        \midrule
        $B_{d}$   & Breathing depth & \SI{12}{mm}\\
        $B_{in}$   & Time for breathing in  & \SI{1500}{ms}\\
        $B_{out}$  & Time for breathing out & \SI{1500}{ms}\\
        $P$  & Pause between breathing out and in & \SI{1000}{ms}\\
        $B_{total}$ & Total time of one breathing cycle & \SI{4000}{ms}\\
        $A$ & The amplitude of  $\cos^{4}$ function & \SI{12}{mm}\\
        $\omega$ & The frequency of $\cos^{4}$ function & $\frac{\pi}{B_{total}}$\\
        \bottomrule
\end{tabular}
\end{table*}

The parameters for the scanner setup are listed in Table~\ref{tab:variablesScanner} and are adaptable to fit most scanners. We selected parameters closely matching those of Gantry 2 at PSI \cite{pedroni2004psi}.
The grid size (\textit{G}) refers to the distance between the spots. The patient positioning table movement is parametrised by its speed (\textit{t$_{T}$}) and settling time (\textit{t$_{Ts}$}). The beam delivery system is characterised by the energy switching time (\textit{t$_{E}$}), the spot transition time (\textit{t$_{Sx}$}, \textit{t$_{Sy}$}) and the dwell time (\textit{t$_{D}$}).

\begin{table*}[h]
\centering
\caption{Proton scanner specific parameters according to the Gantry 2 scanner at the PSI.}\label{tab:variablesScanner}
\begin{tabular}{lll}
    \toprule
    \textbf{Notations} & \textbf{Definition} & \textbf{Variable}\\
    \midrule
      $G$   & Grid size & \SI{3}{mm}\\
      $t_{T}$  & Time for table to move \SI{6}{mm}  & \SI{1000}{ms} \\
      $t_{Ts}$  & Settling time after table movement & \SI{2000}{ms} \\
      $t_{Sx}$   & Spot transition time in $x$ & \SI{3}{ms}  \\
      $t_{Sy}$   & Spot transition time in $y$ & \SI{3}{ms} \\
      $t_{E}$   & Energy switching time & \SI{80}{ms} \\
      $t_{D}$   & Dwell Time & \SI{2}{ms} - \SI{50}{ms} \\
      \bottomrule
\end{tabular} 
\end{table*}

\subsection{Gantry-free and Magnet-free Setups}
We consider five gantry-free or magnet-free options. We investigated two magnet-free delivery options, without a gantry or scanner magnets: the \emph{Stationary Beam} (\emph{SB}) and the \emph{Stationary Variational Beam} (\emph{SVB}).

In the \emph{SB} setup, the tumour passes through the beam through breathing, essentially self-scanning. In the \emph{SVB} setup, the energy is changed at the same time, while the tumour is passing through the beam. 
For the gantry-free setups, we considered three possible scanner heads: the \emph{Scanner XY}, which lacks only the gantry, and the \emph{Scanner X} and \emph{Scanner Y} heads, where one or the other scanner magnet is removed. A summary of the different setups and their respective \emph{DOFs} is shown in Table~\ref{tab:DOFScanner}. 

\begin{table*}[h!]
\centering
\caption{The \emph{DOF} of the beam delivery system in the different setups.}\label{tab:DOFScanner}
\begin{tabular}{l|ccccc} 
\toprule
\textbf{DOF} & \textbf{SB} & \textbf{SVB} & \textbf{Scanner X} & \textbf{Scanner Y} & \textbf{Scanner XY} \\ 
\midrule
Translation $x$ & $-$ & $-$ & $\surd$ & $-$ & $\surd$ \\
Translation $y$ & $-$ & $-$ & $-$ & $\surd$ & $\surd$ \\
Translation $z$ &  $\surd$ &  $\surd$ &  $\surd$ &  $\surd$ &  $\surd$ \\
Rotation & $-$ & $-$ & $-$ & $-$ & $-$ \\
\bottomrule
\end{tabular}
\end{table*}

\begin{figure} [H] 
    \centering
    \begin{subfigure}{0.5\textwidth}
        \centering
        \includegraphics[width=1\linewidth]{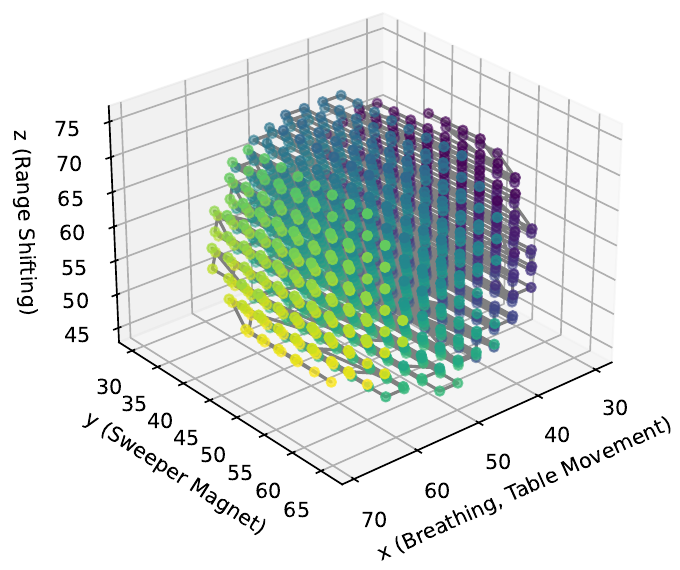}
        \caption{}\label{fig:section_Y}
    \end{subfigure}%
    \hfill 
    \begin{subfigure}{0.5\textwidth}
        \centering
        \includegraphics[width=1\linewidth]{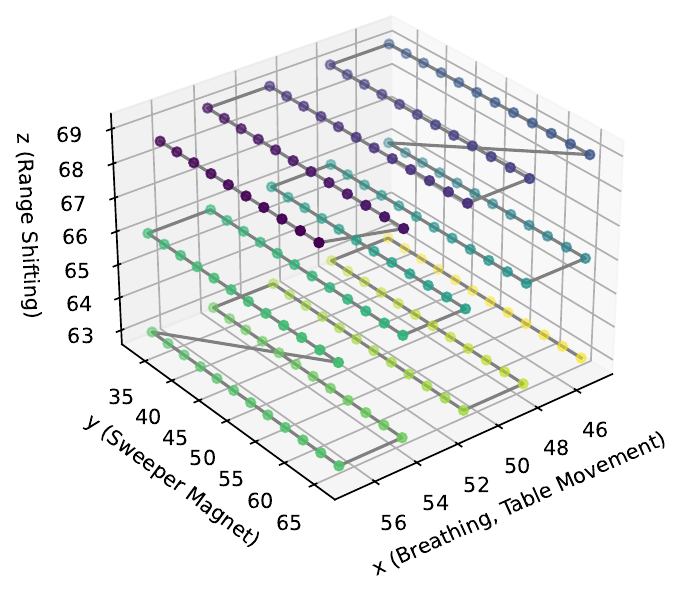}
        \caption{}\label{fig:slice_y}
    \end{subfigure}
    \caption{Schematic showing proposed path within the phantom sphere in a \textit{Scanner Y} setup. (a) sectioned spot positions of the \SI{40}{mm} phantom sphere along the $x$-axis according to breathing depth. (b) The proposed pathway within a section.}
\end{figure}

The scanner setup dictates the order in which spots are scanned, prioritising efficiency by accounting for the translational velocity of each system component. As shown in Table~\ref{tab:variablesScanner}, the table moves the slowest and is therefore the limiting factor in the scanning process, whereas the scanner magnets are the fastest. 
For example, in the \emph{Scanner Y} setup, the target is divided into sections to minimise unnecessary table movement. Within each section, treatment is delivered by the faster components. The size of such a section in the case of the \emph{Scanner Y} setup depends on the breathing depth of the phantom. Figure \ref{fig:section_Y} illustrates such sections within a phantom sphere of \SI{40}{mm} diameter. In each section, we start in the deepest layer along the $z$-axis. While the breathing motion slowly moves the target along the $x$-axis, the scanner magnet quickly moves back and forth along the $y$-axis. Figure~\ref{fig:slice_y} illustrates this movement. When the section is scanned, the table moves the phantom to the next one. It follows this computed trajectory until all points have been visited once. Further details on each delivery system and its respective optimised beam trajectory are provided in Section A of the Supplementary Material.

\subsection{Spot delivery time}
\begin{figure} [H]
    \centering
    \begin{subfigure}{0.55\textwidth}
        \centering
        \includegraphics[width=1\linewidth]{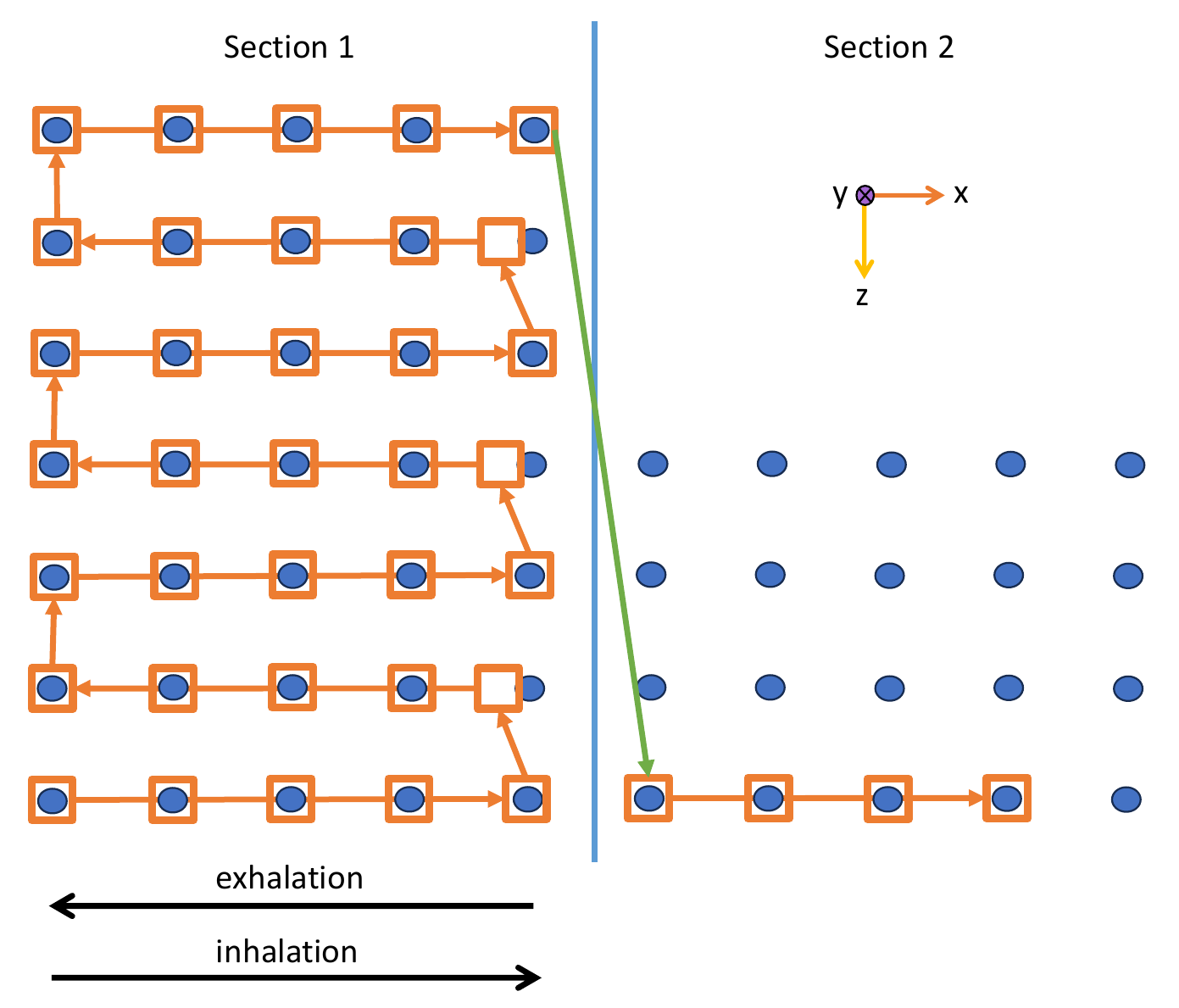}
        \caption{}\label{fig:pathning_SB}
    \end{subfigure}%
    \hfill
    \begin{subfigure}{0.45\textwidth}
        \centering
        \includegraphics[width=1\linewidth]{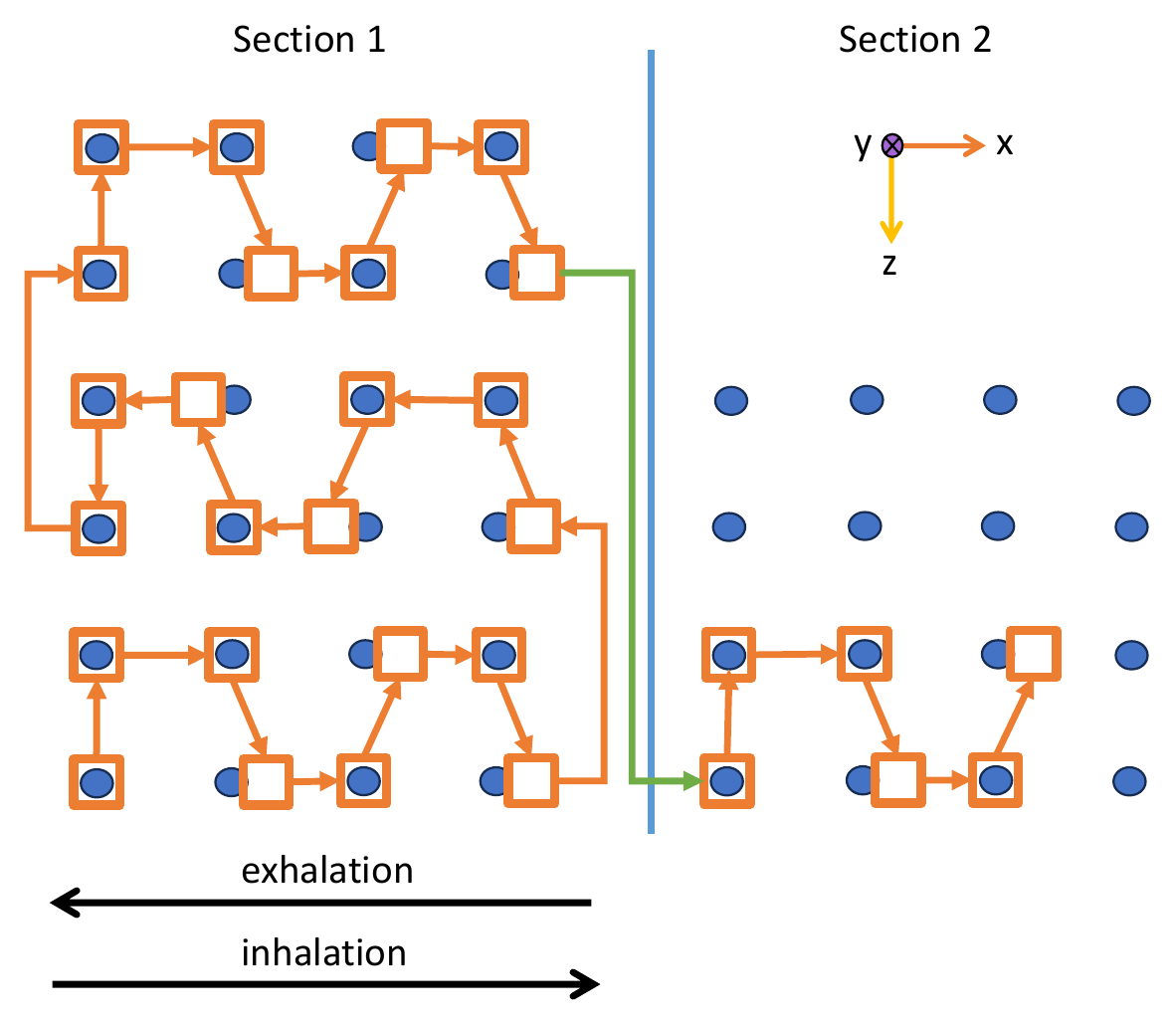}
        \caption{}\label{fig:pathing_SVB}
    \end{subfigure}
    \caption{Schematic showing an example of the correction of beam positions within the phantom in an \emph{SB} and \emph{SVB} setup. The blue points represent the planned positions, while the orange squares show the visited positions. The green arrows represent a table movement to the next section. (a) Shows an example for \emph{SB} scanning mode. (b) Shows an example for \emph{SVB} scanning mode.}\label{fig:pos_corr}
\end{figure}

With the trajectory and the patient's breathing parameters, it is possible to determine the delivery time for each spot. Some positions needed to be corrected if they could not be visited due to time constraints. Whether a position needs to be shifted depends on the breathing velocity and the beam delivery time. If the tumour has already moved more than \SI{0.5}{mm} beyond the planned position, the position is adjusted. In Figure~\ref{fig:pos_corr}, we can see that the change in energy, in combination with the dwell time for beam delivery, causes the position to shift along the breathing direction. In the case of the \emph{SVB}, we can also see that the entire breathing amplitude cannot be used for scanning the tumour. There is insufficient time to properly treat the positions currently located in the beam when the breathing motion inverts from inhalation to exhalation. All scanner modes that rely on the breathing motion to move the tumour through the beam are affected by such corrections.

\subsection{Weight Optimisation}\label{section:WeightOptimisation}
\begin{figure}[H]
    \centering
    \includegraphics[width=1\linewidth]{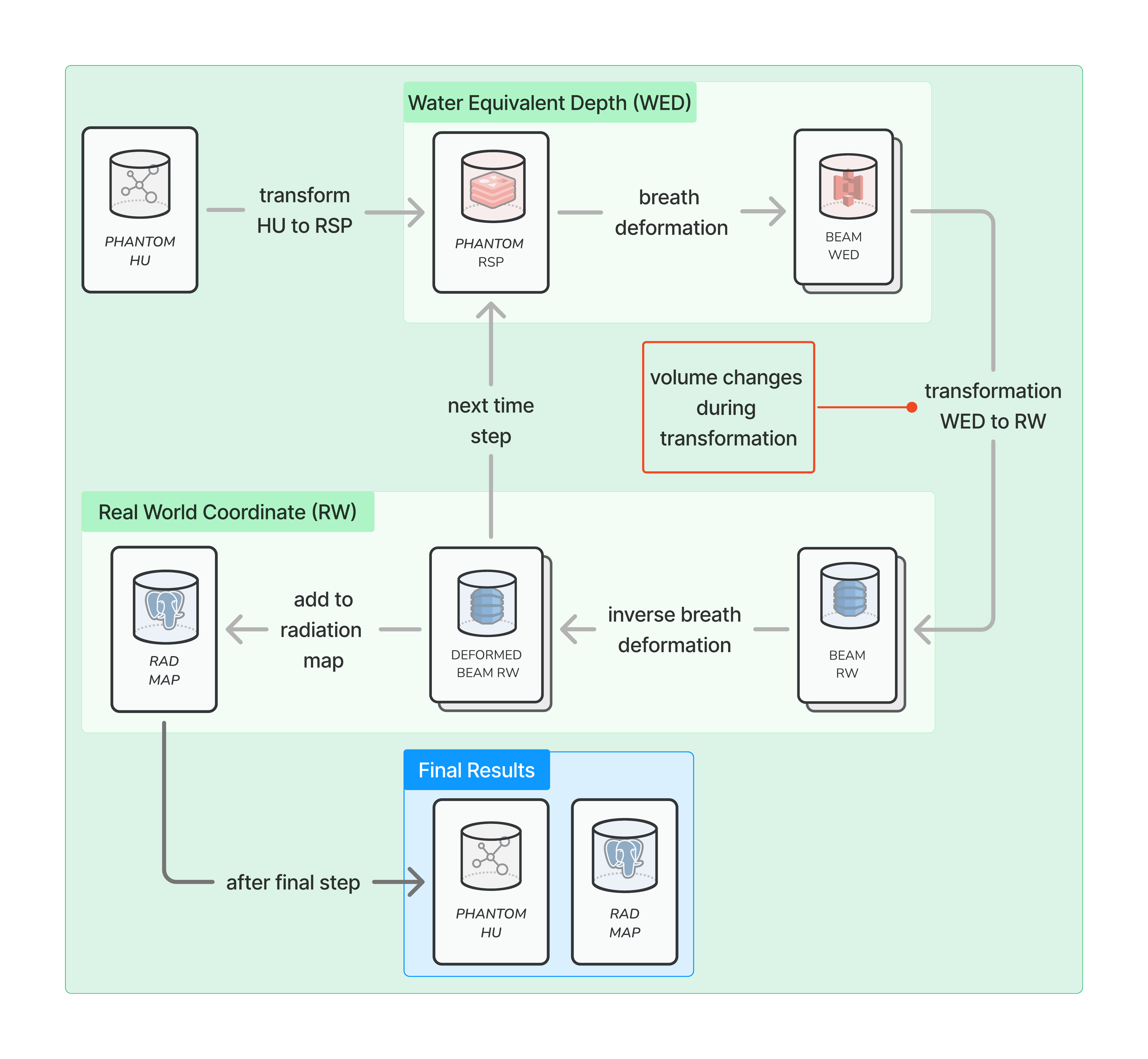}
    \caption{Schematic showing the calculation of the radiation delivered during the treatment. The 3D CT image of the phantom is converted into an RSP map to calculate the beam deformation resulting from density differences. Additionally, the phantom is deformed according to the breathing motion based on the current time step. }\label{fig:protokoll}
\end{figure}

For a complete 4D treatment plan, the beam weights were optimised to ensure homogeneous distribution of the prescribed dose. To guide the optimisation, we evaluated the final distribution using common PT metrics such as $\mathrm{V_{95 \%}}$ and $\mathrm{D_{95 \%}}$. 
Figure~\ref{fig:protokoll} shows a schematic of the treatment simulation and the calculation of the radiation distribution within the phantom. First, the phantom's CT, expressed in Hounsfield Units (HU), is converted to Relative Stopping Power (RSP) \cite{van2018high}. The table used for this transformation is provided in Section B.1 of the Supplementary Material. 
On the transformed CT, we can now apply the breathing deformation at the specific time step $t$. The beam can now be simulated in this deformed map. We approximate the dose distribution of the beam using the following model \cite{schaffner1999dose}:

\begin{eqnarray}\label{eq:Beam}
     D(x; y; z) = N_{p+} \cdot \mathrm{ID}\left(\mathrm{wer}\right) \frac{1}{2 \pi \sigma _{x}^{2} \sigma _{y}^{2}} \cdot \exp\left[ \frac{ - (x_{0} - x)^{2}}{2 \sigma _{x} ^{2}}\right] \cdot \exp\left[ \frac{ - (y_{0} - y)^{2}}{2 \sigma _{y}^{2}}\right]
\end{eqnarray}

Where $N_{p+}$ is the number of particles, ID is the integral dose interpolated from the depth-dose lookup table \cite{rietzel2000correlation} and wer$(x_{0}, y_{0}, z)$ is the \emph{water equivalent range} along the central axis of the beam and $\sigma _{x}$, $\sigma _{y}$ are the standard deviations of the Gaussians in $x$- and $y$- directions. We extracted the tables for $\sigma_{x}$, $\sigma_{y}$, and ID from a Monte Carlo simulation of PSI's Gantry 2 for a beam in wer at 170 MeV. The integral dose and standard deviation obtained from this beam are provided in Section B.2 in the Supplementary Material.

Because the radiation dose is calculated using a water-equivalent range (WER), the dose distribution must be corrected for the target material's density, which involves transforming the beam from a water-equivalent depth (WED) coordinate system to a real-world (RW) coordinate system. During this step, it is essential to account for the volume change of each voxel. To obtain the radiation distribution during the phantom's resting phase, the radiation map is inverse-transformed using the previously applied breathing deformation. 
These steps are repeated for all the pencil beams delivered at a position ($x$, $y$, $z$) at the respective time step $t$. The calculated radiation doses are summed to yield the final radiation distribution in the phantom after treatment. The radiation is measured in Gray (Gy).

To optimise the weights to generate a homogeneous radiation field, we first define constraints on the final radiation distribution. We optimise the weights by minimising the loss function in Equation~\ref{eq:Loss} using the Adam optimiser \cite{kingma2014adam}. The metrics and constraints used in the optimisation are reported in Table~\ref{tab:metricsconstraints}. The optimal values and constraints defined for these metrics are reported in Table~\ref{tab:todo}.

\begin{table*}[h]
\centering
\caption{Metrics and Constraints used for the optimisation of weights for the dose distribution.} \label{tab:metricsconstraints}
    \begin{tabular}{lll}
    \toprule
    \textbf{Notations} & \textbf{Units} & \textbf{Definition}\\
    \midrule
    $\mathrm{V_{95 \%}}$ & [\%] & Volume that receives at least 95\% of the prescribed dose. \\
    $\mathrm{D_{95 \%}}$ & [Gy] & The minimum dose received by at least 95\% of the volume. \\ 
    $\mathrm{D_{max}TV}$ & [\%] & The maximal dose as a percentage of the prescribed dose. \\ 
    $\mathrm{D_{max}HT}$ & [Gy] & The maximal dose encountered in the healthy tissue. \\
    \bottomrule
    \end{tabular}
\end{table*}

\begin{table*}[h]
\centering
\caption{Used optimisation metrics.}\label{tab:todo}
    \begin{tabular}{llc}
        \toprule
        \textbf{Notations} & \textbf{Definition} & \textbf{Optimisation Metric}\\
        \midrule
        $\mathrm{L}_{V}$ & The ideal value for $\text{V}_{95\%}$ &\SI{98}{\%} \\ 
        $\mathrm{L}_{D}$ & The ideal value for $\text{D}_{95\%}$ &\SI{10}{Gy}\\ 
        $\mathrm{L}_{max}\mathrm{TV}$ & The maximum for $\text{D}_{max}\text{TV}$ & $<$ \SI{120}{\%}\\ 
        $\mathrm{L}_{max}\mathrm{HT}$ & The maximum for $\text{D}_{max}\text{HT}$ & $<$ \SI{7}{Gy}\\ 
        \bottomrule
    \end{tabular}
\end{table*}

Equation~\ref{eq:Loss} shows the loss function. 

\begin{equation}\label{eq:Loss}
    \mathcal{L} =\mathcal L_V + \mathcal L_D + c_{max} \left( \mathcal L_{DmaxTV} + \mathcal L_{DmaxHT} \right), 
\end{equation}
with $c_{max}$ as a weight, determining the importance of $\mathcal L_{DmaxTV}$ and $\mathcal L_{DmaxHT}$. The different components of the loss function can be seen in Equation~\ref{eq:Loss_V} to \ref{eq:Loss_maxHT}.

\begin{equation}\label{eq:Loss_V}
    \mathcal L_V = \max\{\mathrm{L}_V - \mathrm{V_{95\%}}, 0\}
\end{equation}
\begin{equation}\label{eq:Loss_D}
    \mathcal L_D = \max\{\mathrm{L}_D-\mathrm{D_{95\%}}, 0\}
\end{equation}
\begin{equation}\label{eq:Loss_maxTV}
    \mathcal L_{DmaxTV} = \max\{\mathrm{D}_{max}\mathrm{TV} - \mathrm{L}_{max}\mathrm{TV}, 0\}
\end{equation}
\begin{equation}\label{eq:Loss_maxHT}
    \mathcal L_{DmaxHT} = \mathrm{D}_{max} \mathrm{HT} - \mathrm{L}_{max} \mathrm{HT}
\end{equation}

To assess the final dose conformity, the following metrics were calculated: 
$\mathrm{V_{98 \%}}$, $\mathrm{V_{100 \%}}$, $\mathrm{D_{2 \%}}$, $\mathrm{D_{50 \%}}$, $\mathrm{D_{98 \%}}$ of the target volume. To assess the homogeneity of the radiation distribution, the homogeneity index (HI) \cite{kataria_homogeneity_2012} and the conformation number (CN) \cite{van1997conformation} were calculated. The homogeneity index was calculated with the following equation:
\begin{equation}
   \mathrm{ HI = \frac{D_{2\%} - D_{98\%}}{D_{50\%}}, }
\end{equation}
where a value close to zero indicates high homogeneity within the target volume. The conformation number was calculated with the equation:
\begin{equation}
   \mathrm{ CN = \frac{TV_{RI}}{TV} \cdot \frac{TV_{RI}}{V_{RI}}, }
\end{equation}
where TV is the target volume, $\mathrm{TV_{RI}}$ is the target volume covered by 95\% of the isodose and $\mathrm{V_{RI}}$ is the total volume covered by 95\% of the isodose. A value close to one represents high homogeneity. 

\newpage

\section{Results}
We calculated the delivery time for the different phantom sizes, setups, and dwell times in Section~\ref{section:DeliveryTime}. We show a potential radiation delivery distribution of an optimised delivery plan for the \emph{SB}, the \emph{SVB} and the \emph{Scanner XY} delivery systems in Section~\ref{section:DoseDistribution}. The delivery plan is illustrated with both the planned and irregular breathing patterns to demonstrate what happens when breathing is not synchronised with the treatment plan.

\subsection{Delivery Time}\label{section:DeliveryTime}
\begin{figure}[H]
    \centering
    \begin{subfigure}{0.61\textwidth}
        \centering
        \includegraphics[width=\textwidth]{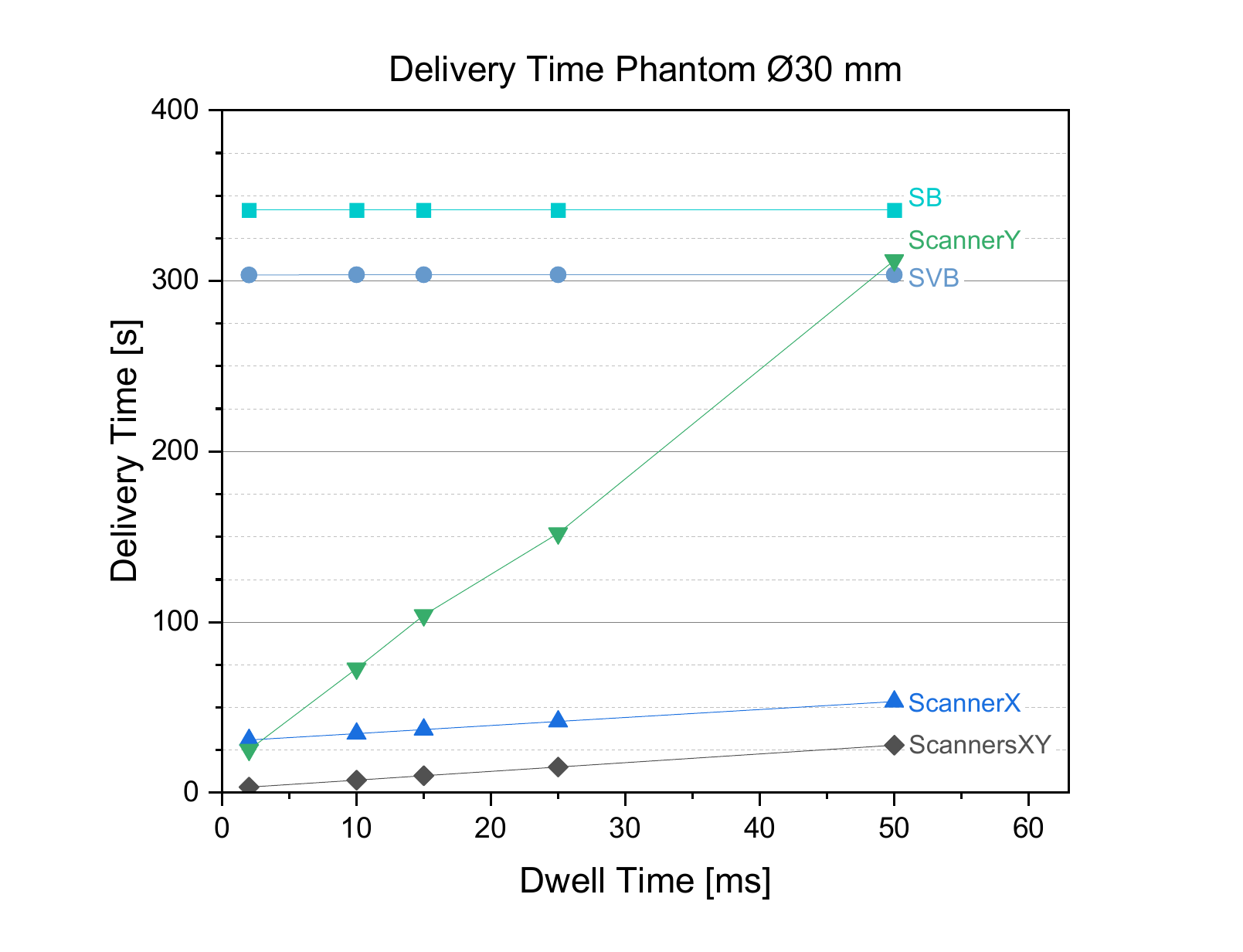}
        \caption{}\label{fig:DwellTime30mm}
    \end{subfigure}%
    \hfill
    \begin{subfigure}{0.61\textwidth}
        \centering
        \includegraphics[width=\textwidth]{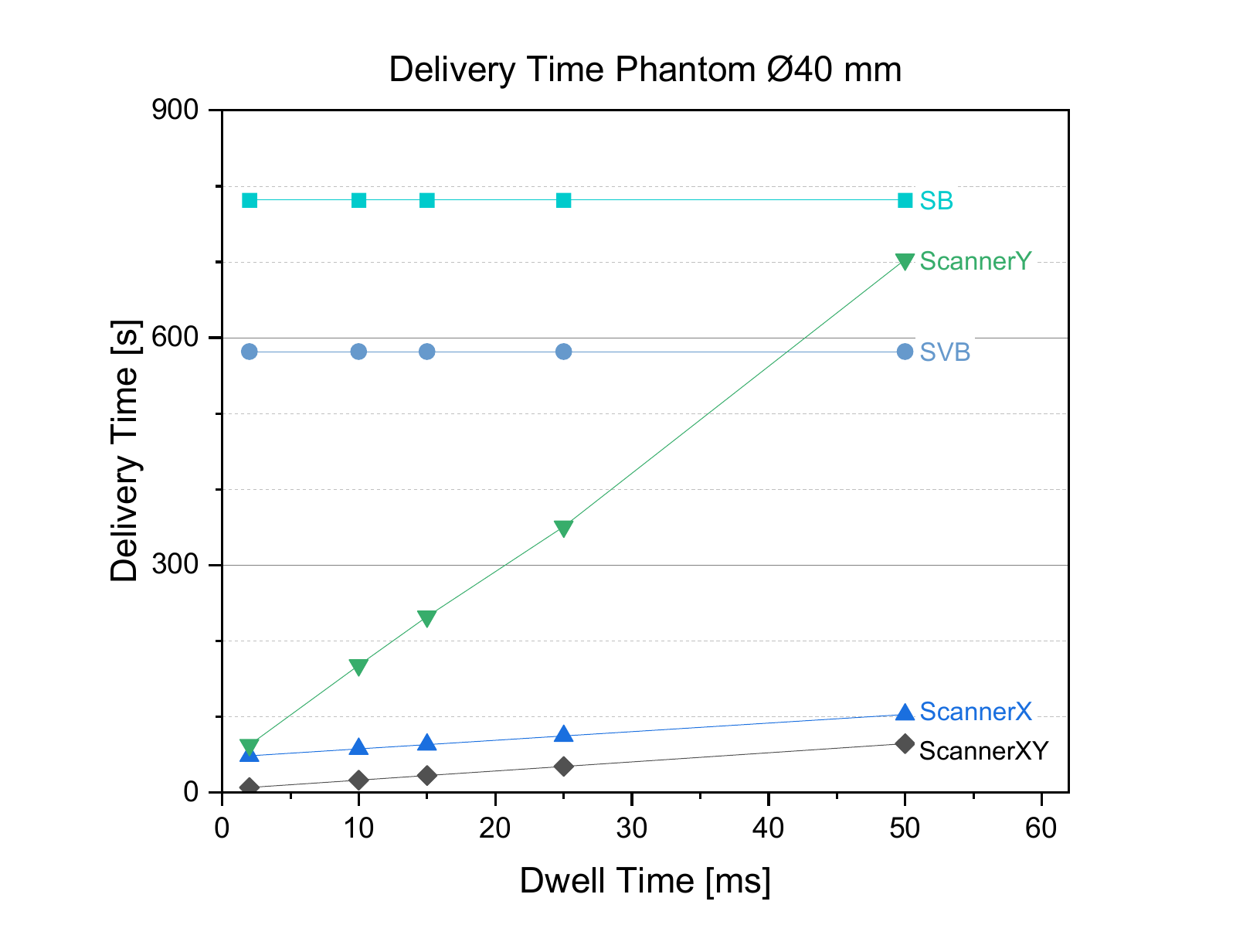}
        \caption{}\label{fig:DwellTime40mm}
    \end{subfigure}
    \hfill 
    \begin{subfigure}{0.61\textwidth}
        \centering
        \includegraphics[width=\textwidth]{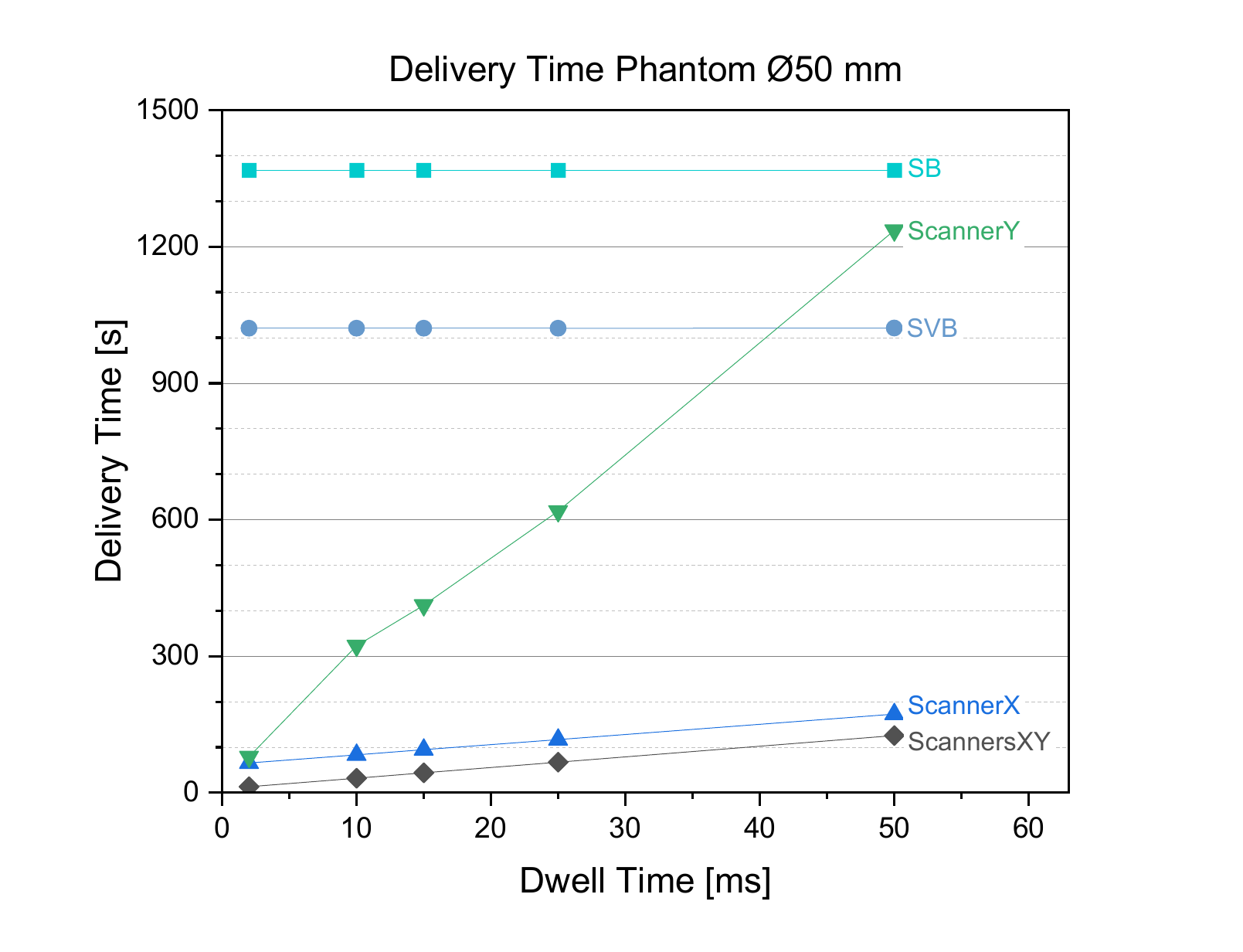}
        \caption{}\label{fig:DwellTime50mm}
    \end{subfigure}
    \caption{Calculated delivery time for the phantom spheres with diameter \SI{30}{mm} Figure~(a), \SI{40}{mm} Figure~(b) and \SI{50}{mm} Figure~(c). Each delivery mode and set of different dwell times, assuming regular breathing without interruption.}\label{fig:time_diag}
\end{figure}

The calculated delivery time is shown in Figure~\ref{fig:time_diag} for all dwell times and tumour sizes. The delivery time was calculated assuming uninterrupted, regular breathing, with the parameters listed in Section~\ref{section:TreatmentSimulation} Table~\ref{tab:PhantomVariables}.

The results show that \emph{SB} and \emph{SVB} setups, with no gantry and no scanner magnets, are the slowest overall, with a duration of approximately \SI{6}{min} and \SI{5}{min} for the smallest and around \SI{23}{min} and \SI{17}{min} for the largest spherical tumour, respectively. As these setups depend entirely on the breathing motion and the table's movement during tumour scanning, the dwell time does not affect the total delivery time in our current setup.
The fastest scanner mode is the \emph{Scanner XY} setup, which uses both the $x$ and $y$ scanner magnets to account for breathing motion, thus eliminating the need for table movement to deliver a conformal dose. In this case, the delivery time highly depends on the dwell time. For a dwell time of \SI{2}{ms} and the smallest target volume, the delivery time is approximately \SI{3}{s}. For a dwell time of \SI{50}{ms}, the delivery can take up to \SI{28}{s}. For larger tumours, the time increases to approximately \SI{2}{min} for the largest tumour and a dwell time of \SI{50}{ms}. For the \emph{ScannerX} delivery setup, the delivery time increases linearly with both tumour size and dwell time, with the same rate as the \emph{Scanner XY} setup. For the \emph{ScannerY} setup, the time increases considerably with dwell time. For a dwell time of \SI{50}{ms}, this setup is even slower than the \emph{SVB} delivery mode.

\subsection{Optimised Dose Distribution} \label{section:DoseDistribution}
The dose distribution was optimised for three of the five setups: \emph{SB}, \emph{SVB}, and \emph{Scanners XY}, with a dwell time of \SI{10}{ms}.  The setup \emph{SB} and \emph{SVB} have a unique pathing and beam positioning, and the setup \emph{Scanners XY} is the most time-efficient. The optimisation was performed for each target volume size. With the optimised weight, the plans were then tested under irregular breathing conditions. An example of an irregular breathing pattern is provided in Section B.3 in the Supplementary Material.

\begin{figure}[H]
    \centering
    \begin{subfigure}{0.61\textwidth}
        \centering
        \includegraphics[width=\textwidth]{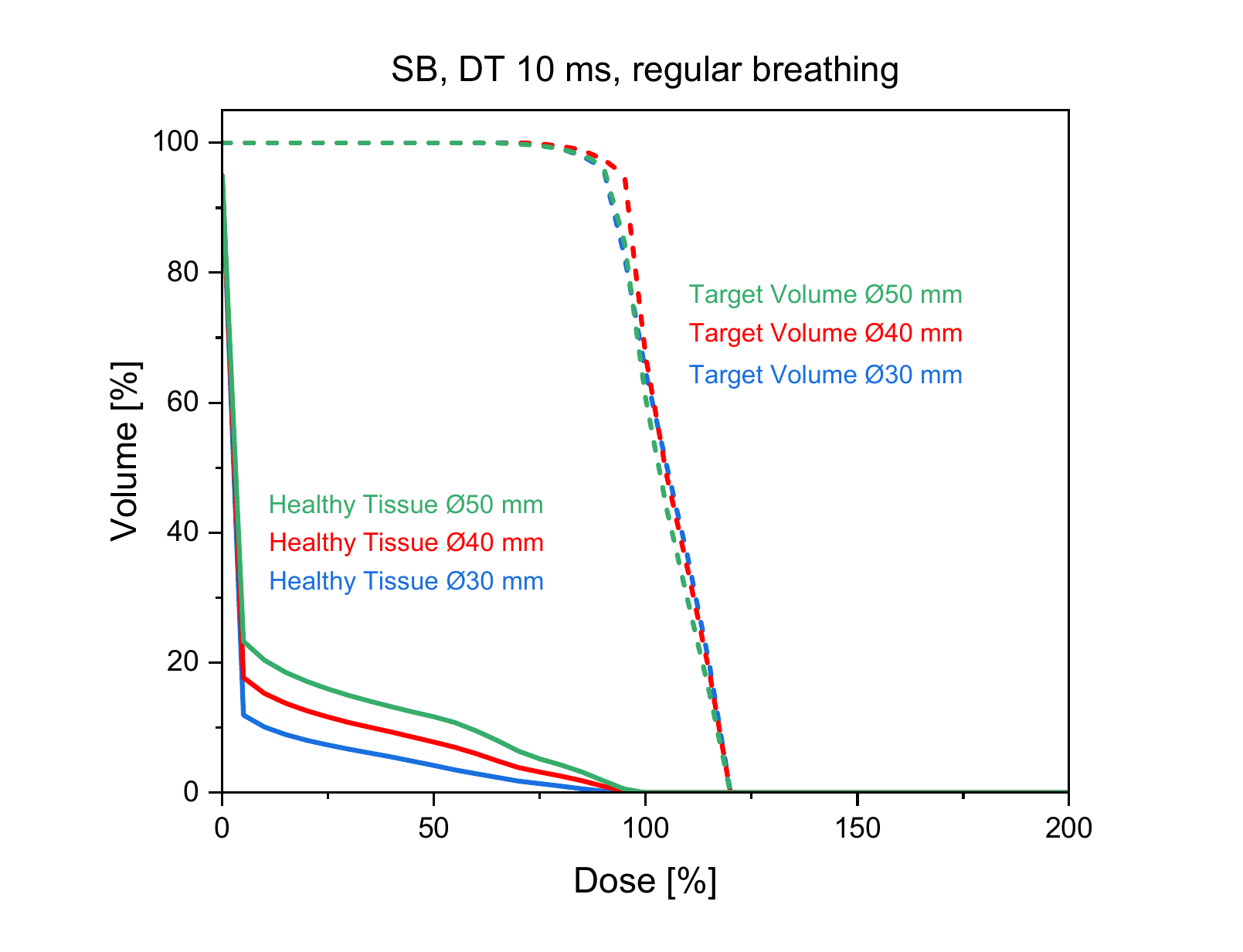}
        \label{fig:DVH_SB_reg}
    \end{subfigure}
    \vspace{1em} 
    \begin{subfigure}{0.61\textwidth}
        \centering
        \includegraphics[width=\textwidth]{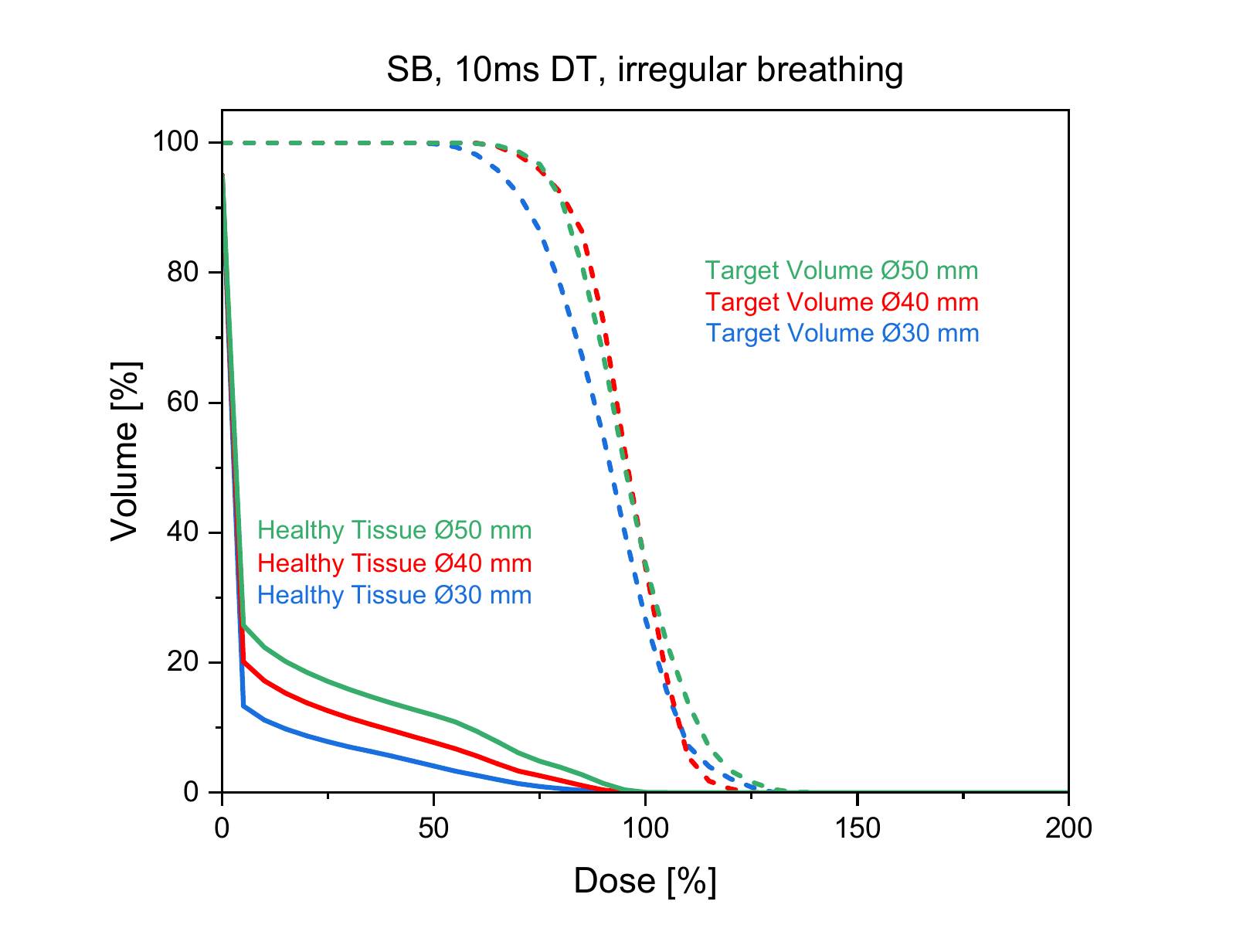}
        \label{fig:DVH_SB_irreg}
    \end{subfigure}
    \caption{DVH of the \emph{SB} scanner mode, for regular breathing and irregular breathing. (Red) DVH for the \SI{30}{mm} diameter phantom. (Green) DVH for the \SI{40}{mm} diameter phantom. (Blue) DVH for phantom of \SI{50}{mm} diameter. The respective V$_{95 \%}$, D$_{95 \%}$ and D$_{max}$TV values for regular and irregular breathing are provided in Section C in the Supplementary Material.}\label{fig:DVH_SB}
\end{figure}

\begin{figure}[H]
    \centering
     \includegraphics[width=\textwidth]{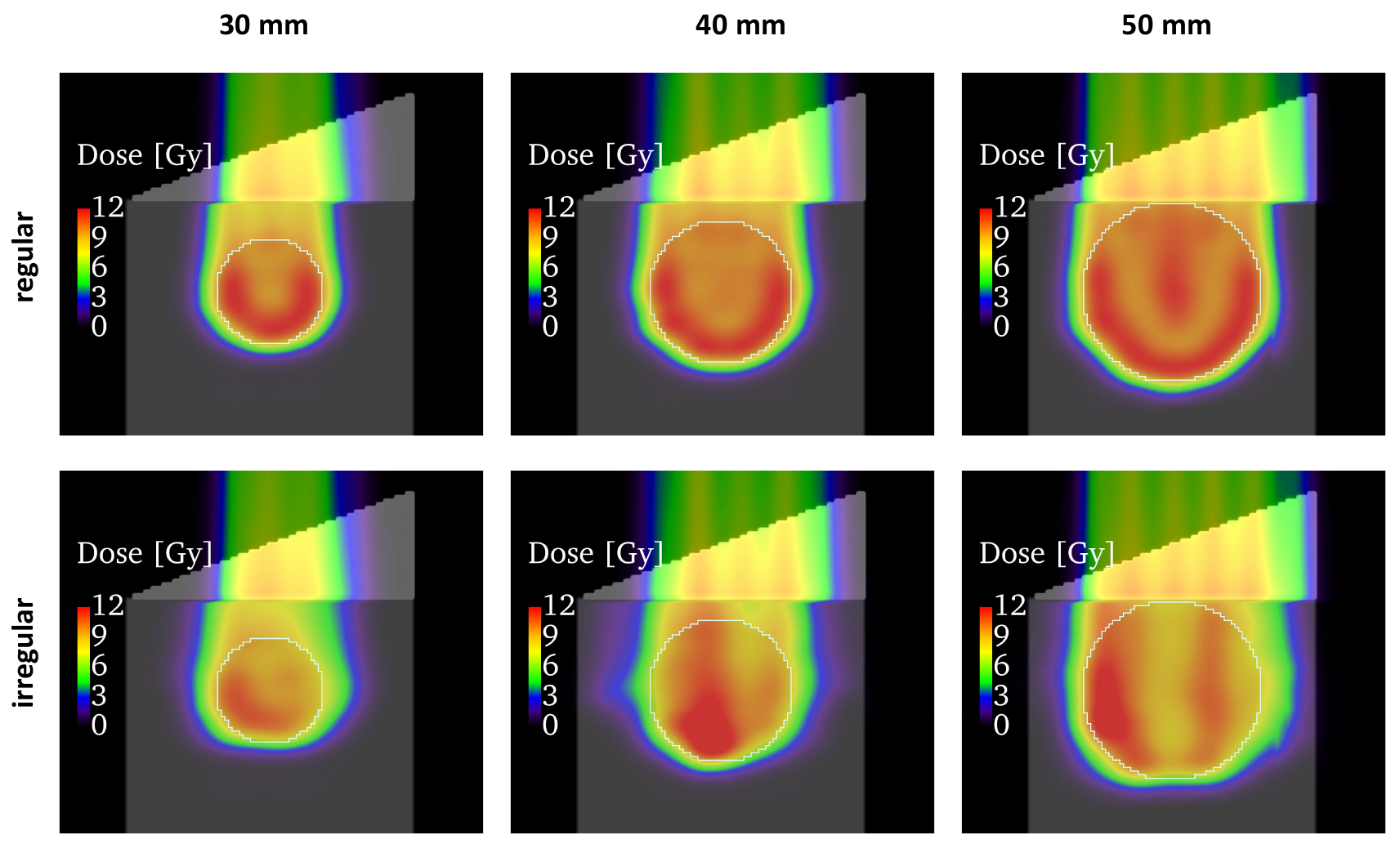}
    \caption{Radiation map for \emph{SVB} delivery system, all three phantom diameters. The top row of maps shows radiation distribution for regular and predicted breathing. The bottom row illustrates the outcome of the delivery plan under non-predicted breathing conditions. }\label{fig:RadMap_SVB}
\end{figure}

\begin{figure}[H]
    \centering
    \begin{subfigure}{0.61\textwidth}
        \centering
        \includegraphics[width=\textwidth]{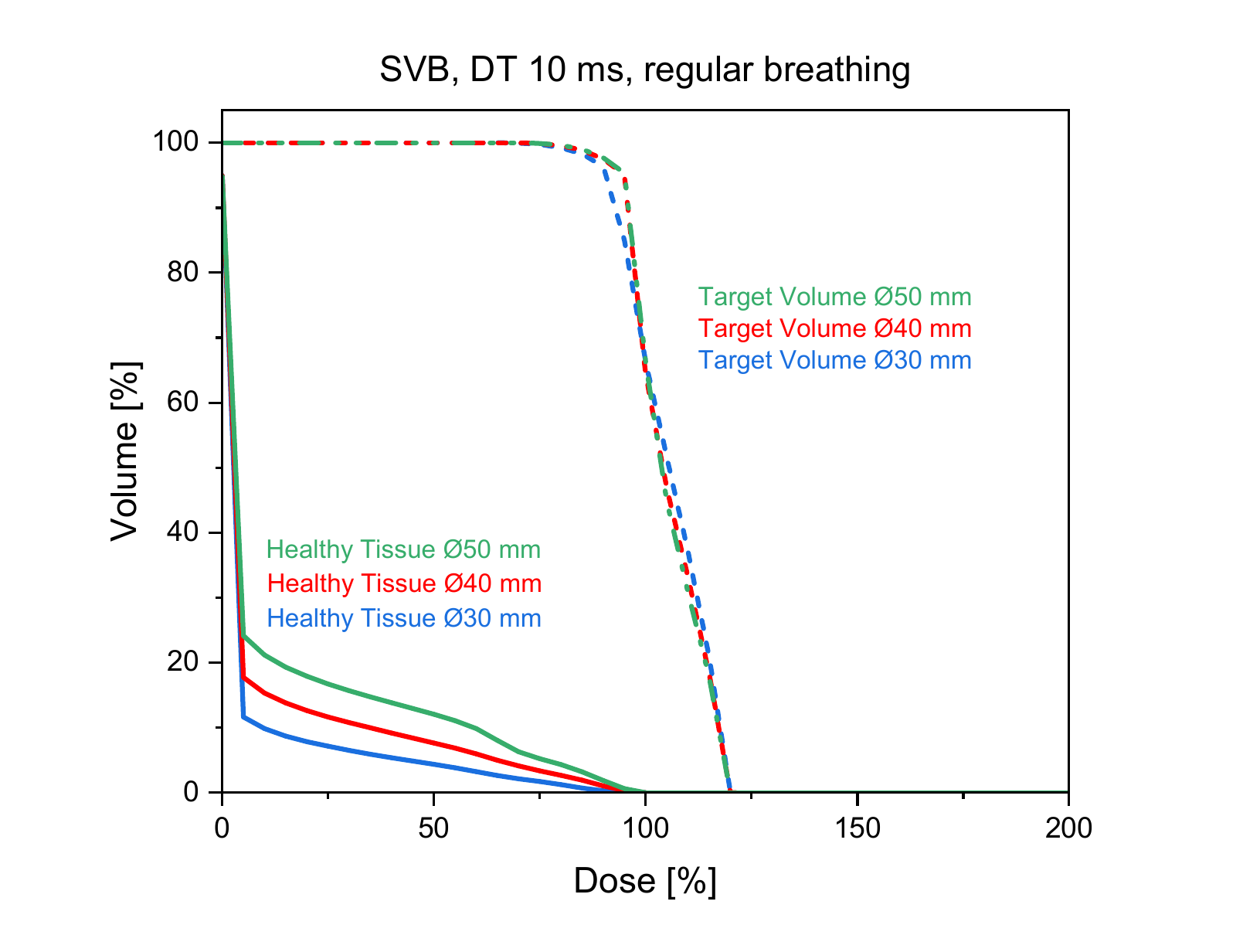}
        \label{fig:DVH_SVB_reg}
    \end{subfigure}
    \vspace{1em} 
    \begin{subfigure}{0.61\textwidth}
        \centering
        \includegraphics[width=\textwidth]{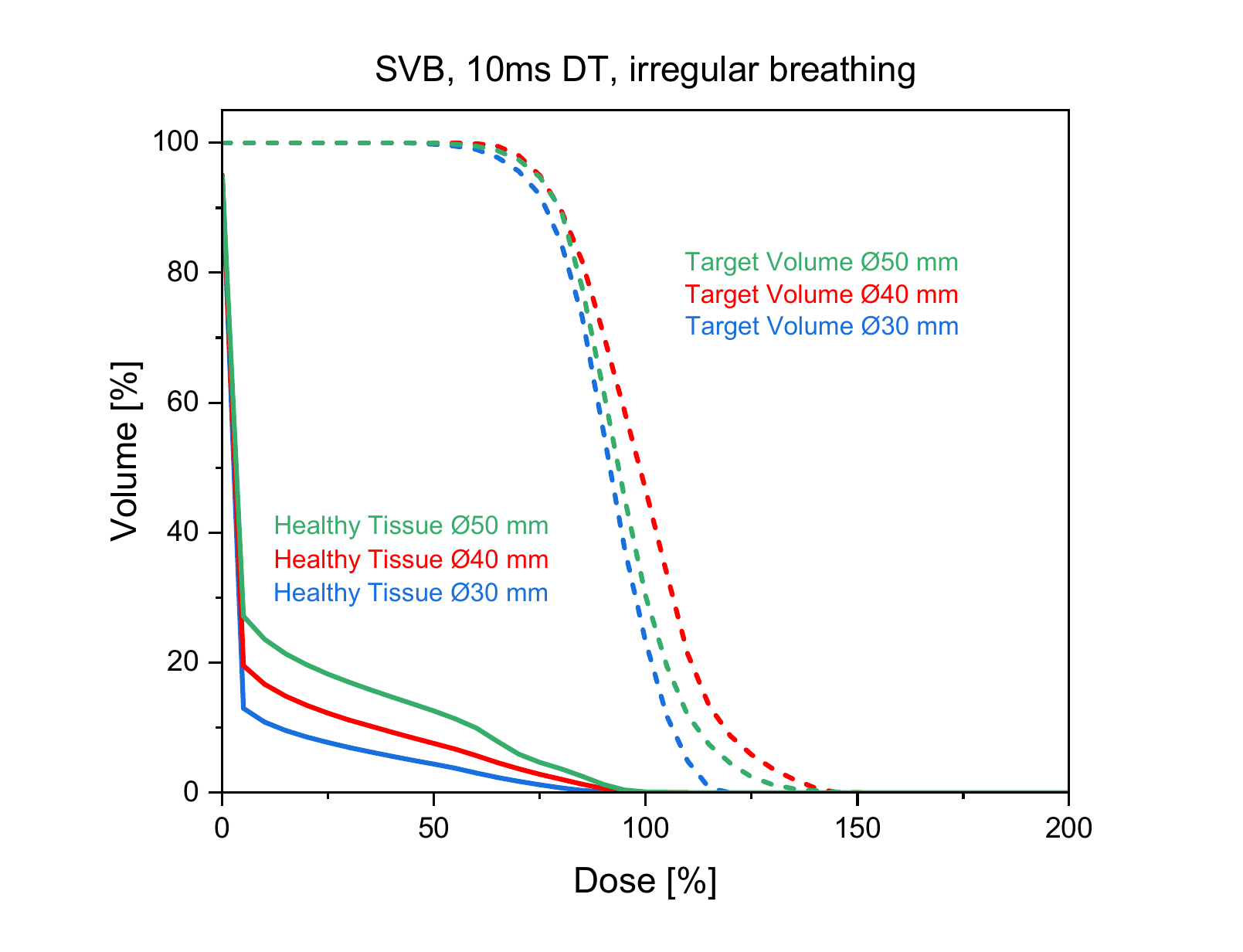}
        \label{fig:DVH_SVB_irreg}
    \end{subfigure}
    \caption{DVH of the \emph{SVB} scanner mode, for regular breathing and irregular breathing. (Red) DVH for the \SI{30}{mm} diameter phantom. (Green) DVH for the \SI{40}{mm} diameter phantom. (Blue) DVH for phantom of \SI{50}{mm} diameter. The respective V$_{95 \%}$, D$_{95 \%}$ and D$_{max}$TV values for regular and irregular breathing are provided in Section C in the Supplementary Material.}\label{fig:DVH_SVB}
\end{figure}

\begin{figure}[H]
    \centering
     \includegraphics[width=\textwidth]{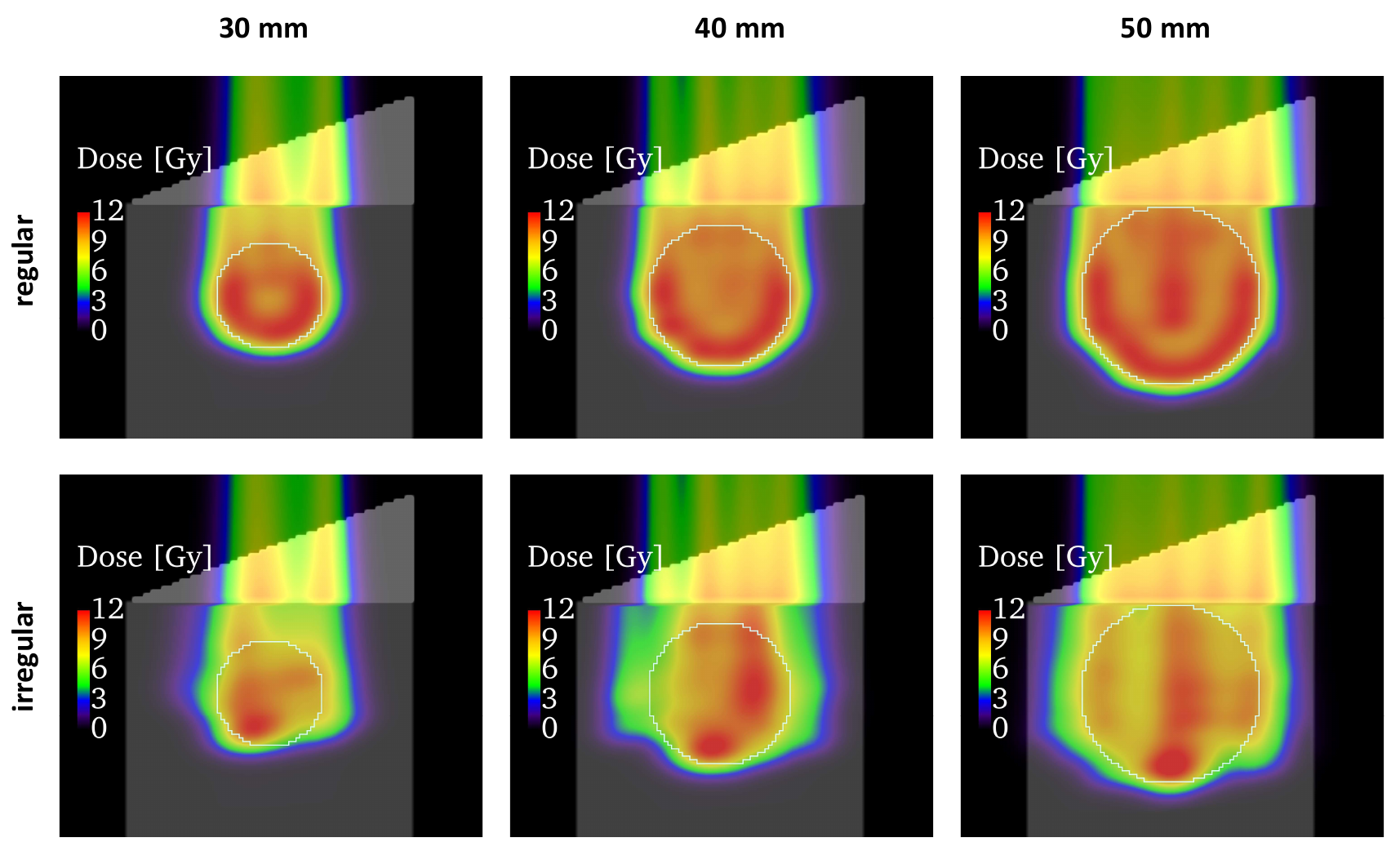}
    \caption{Radiation map for \emph{Scanners XY} delivery system, all three phantom diameters. The top row of maps shows radiation distribution for regular and predicted breathing. The bottom row illustrates the outcome of the delivery plan under non-predicted breathing conditions. }\label{fig:RadMap_ScannersXY}
\end{figure}

\begin{figure}[H]
    \centering
    \begin{subfigure}{0.61\textwidth}
        \centering
        \includegraphics[width=\textwidth]{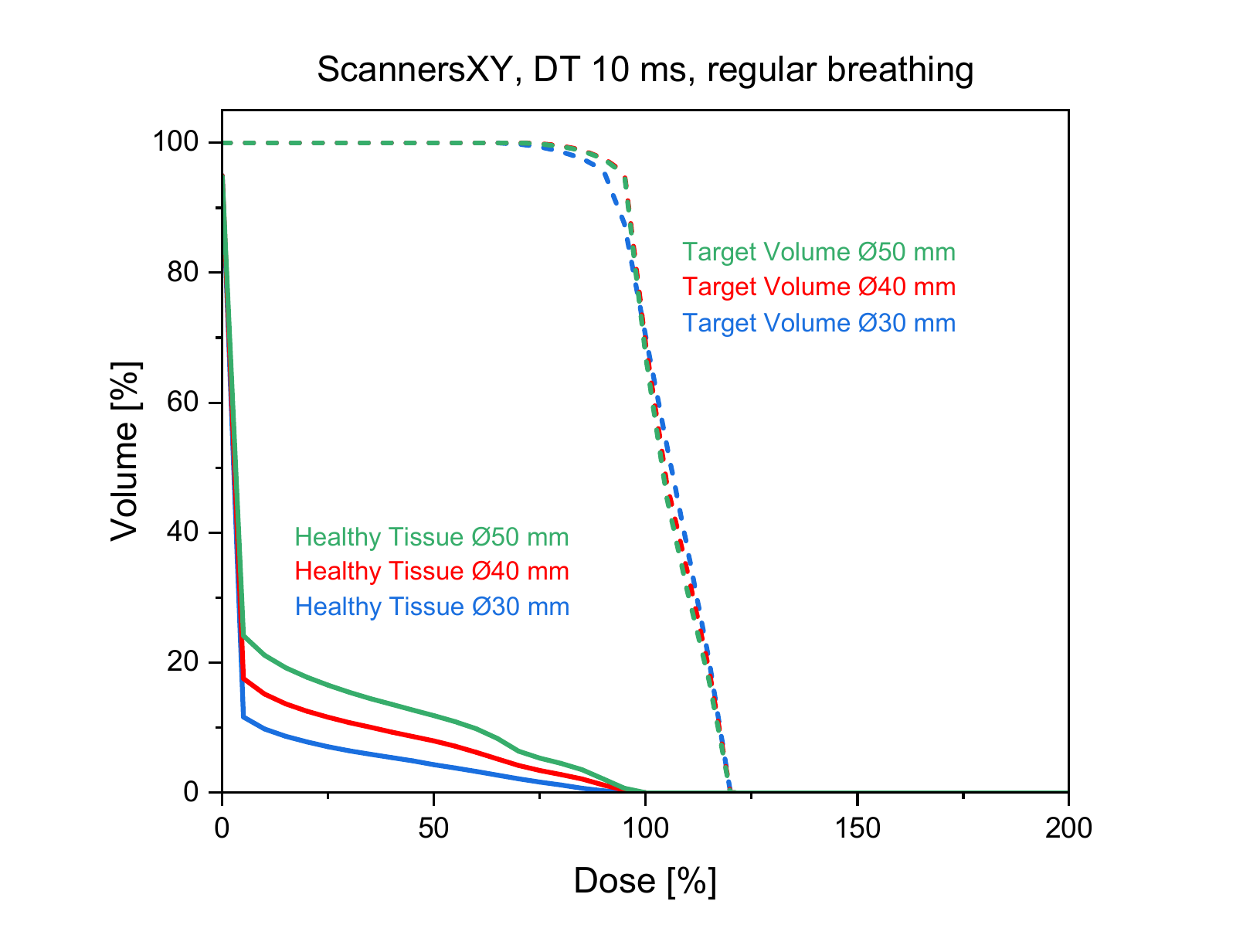}
        \label{fig:DVH_ScannersXY_reg}
    \end{subfigure}
    \vspace{1em} 
    \begin{subfigure}{0.61\textwidth}
        \centering
        \includegraphics[width=\textwidth]{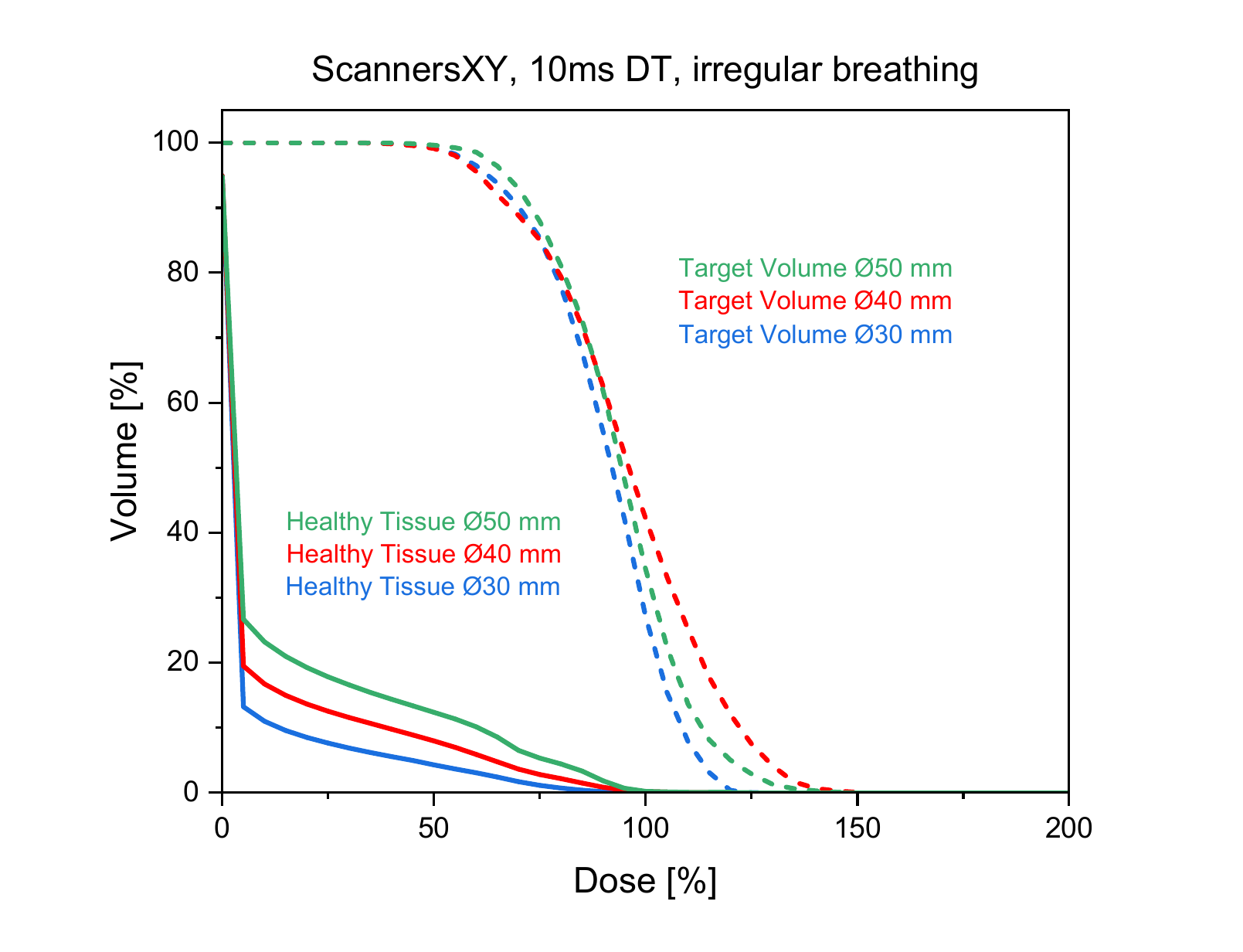}
        \label{fig:DVH_ScannersXY_irreg}
    \end{subfigure}
    \caption{DVH of the \emph{SB} scanner mode, for regular breathing and irregular breathing. (Red) DVH for the \SI{30}{mm} diameter phantom. (Green) DVH for the \SI{40}{mm} diameter phantom. (Blue) DVH for phantom of \SI{50}{mm} diameter. The respective V$_{95 \%}$, D$_{95 \%}$ and D$_{max}$TV values for regular and irregular breathing are provided in Section C in the Supplementary Material.}\label{fig:DVH_ScannersXY}
\end{figure}

In Figure~\ref{fig:RadMap_SB}, \ref{fig:RadMap_SVB} and \ref{fig:RadMap_ScannersXY}, we can see the resulting radiation maps for the scanner mode \emph{SB}, \emph{SVB} and \emph{Scanner XY} respectively. It displays the maps for all three target volume sizes under regular and irregular breathing conditions. Figure~\ref{fig:DVH_SB}, \ref{fig:DVH_SVB}, and \ref{fig:DVH_ScannersXY} show the respective dose-volume histograms (DVH) for each scanner mode.  All the parameters used for the optimisation are provided in Section B.3 in the Supplementary Material, along with the metrics used to evaluate the final radiation map. 

\begin{figure}[H]
    \centering
    \includegraphics[width=\textwidth]{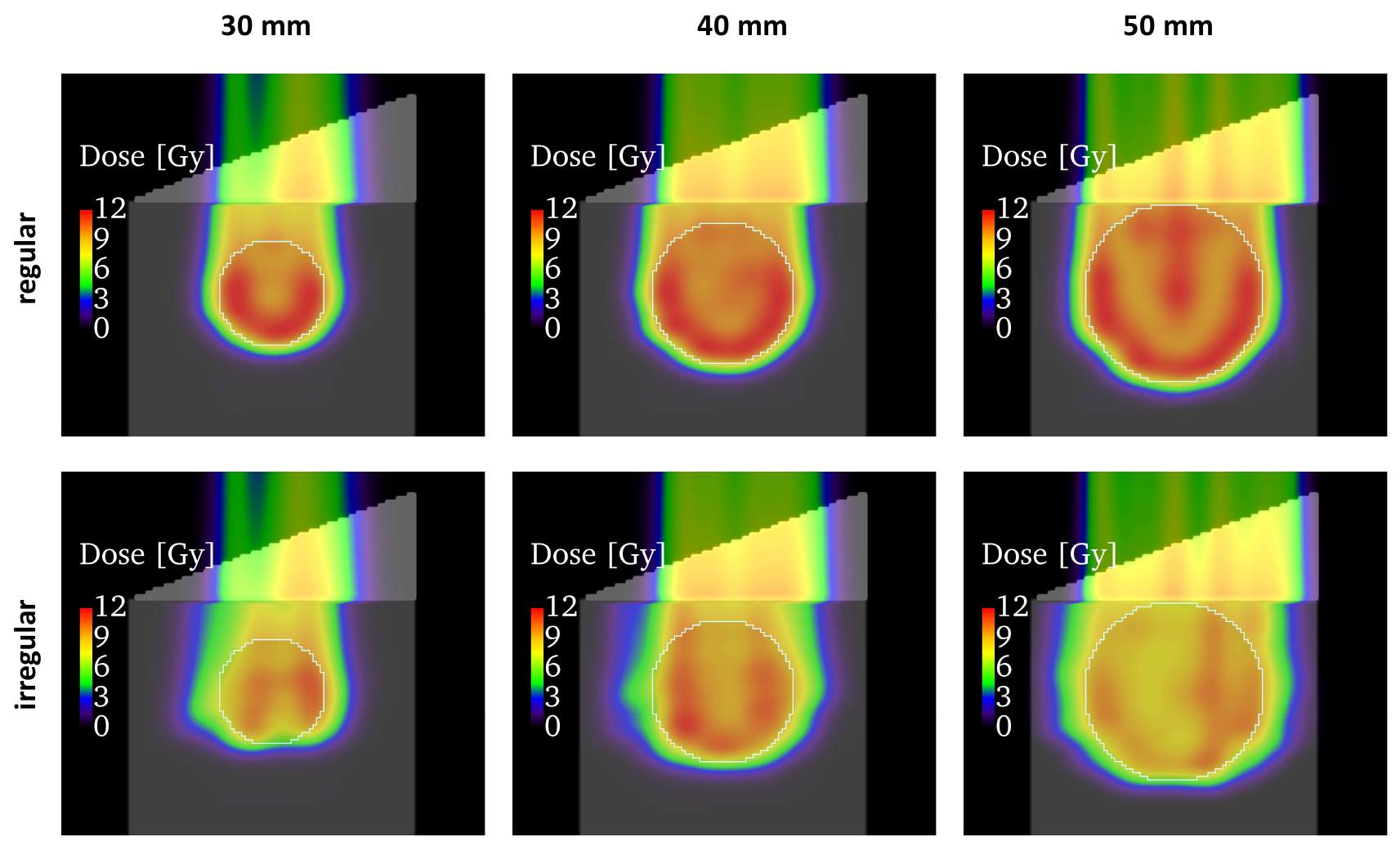}
    \caption{Radiation map for \emph{SB} delivery system, all three phantom diameters. The top row of maps shows radiation distribution for regular and predicted breathing. The bottom row illustrates the outcome of the delivery plan under non-predicted breathing conditions. }\label{fig:RadMap_SB}
\end{figure}

\section{Disscusion}
This paper presents a 4D pencil beam delivery optimisation, a novel motion management approach for proton radiotherapy. 
This tool enables the planning of treatment for moving targets such as liver tumours, under the condition of a known or predicted breathing pattern. The tool was tested on a 4D phantom, representing a liver tumour under respiration. Furthermore, we evaluated this tool for four different delivery system setups and five different delivery trajectories, including a completely magnet-free scanner head. A summary of the different trajectories and their respective delivery setups is reported in Table~\ref{tab:DOFScanner}. We tested the various setups and trajectories with spherical target volumes of three diameter sizes: \SI{30}{mm}, \SI{40}{mm} and \SI{50}{mm} and with different dwell times from \SI{2}{ms} to \SI{50}{ms}. This tool enabled us to determine potential delivery times for the various setups. Additionally, we were able to simulate the possible dose distribution according to the plan and evaluate its robustness to irregular breathing.

\subsection{Delivery Time}
We calculated the delivery time to evaluate the planned trajectories and system setups (see Figure~\ref{fig:time_diag}). We can see that the \emph{Scanner XY} scanner head is the fastest. This setup is not dependent on either breathing or table motion to move the target volume through the beam. Instead, the setup uses scanner magnets for the translational \emph{DOF}. The next fastest is the \emph{Scanner X} setup, which features a scanner magnet aligned along the breathing direction, or the $x$-axis. It is slower than the \emph{Scanner XY} because it needs to wait for the table to move the target volume along the $y$-axis.

The treatment duration for \emph{Scanner Y} scanner head depends heavily on the dwell time. With a dwell time of \SI{50}{ms}, the delivery time is longer than for the \emph{SVB} setup. This additional time is due to the setup's dependence on breathing motion to move the target volume through the beam. When the target volume is larger than the breathing depth, meaning the breathing motion is insufficient to move the entire target volume through the beam, additional table motion is required to move the target along the $x$-axis. Furthermore, the increase in dwell time reduces the number of positions that can be visited during a single breathing cycle. For a dwell time of \SI{2}{ms}, it is possible to treat 12 positions along the $y$-axis. For \SI{25}{ms}, it is only possible to treat two sites before the tumour has moved too far from the planned position. With a dwell time of \SI{50}{ms}, only one position can be treated and is thus less efficient than the \emph{SVB} setup.

The \emph{SB} and \emph{SVB} setups are generally the slowest. These setups rely on the patient's breathing and the table's motion to move the target volume. Both cases require adjustments to some planned positions, as the delivered locations exceed our set difference threshold of \SI{0.5}{mm}. The delivery time of \emph{SB} and \emph{SVB} setups is not affected by the dwell time, as the breathing velocity is much slower.  

\subsection{Dose Distribution}
To further evaluate our planning tool and a magnet-free delivery, we simulated the delivery of the treatment plans. These allowed us to assess how shifting positions in the plans for the \emph{SB} and \emph{SVB} setups affects the overall dose distribution, and how a gantry-free setup, such as the \emph{Scanner XY}, affects the radiation distribution in healthy tissues. Furthermore, we tested the treatment plan under irregular breathing conditions to evaluate its robustness. 

We can observe in Figures~\ref {fig:RadMap_SB} and \ref{fig:RadMap_SVB}, for setups \emph{SB} and \emph{SVB}, that the correction of the positions does not seem to affect the dose distribution when compared to the setup \emph{Scanners XY} in Figure~\ref{fig:RadMap_ScannersXY}. The metrics calculated from the radiation maps provided in Section C in the Supplementary Material further support this. The DVHs in Figures~\ref{fig:DVH_SB}, \ref{fig:DVH_SVB} and \ref{fig:DVH_ScannersXY}, for the setups \emph{SB}, \emph{SVB}, and \emph{Scanners XY}, we can see that the distribution of radiation within the target volume is very similar across all three target sizes. 
However, as the size increases, more healthy tissue is affected by radiation. In the respective radiation maps, it is even possible to see the stationary beams' position, as the dose is higher in those regions of healthy tissue.

Even though the radiation dose in healthy tissue may be within an acceptable range for smaller tumours, it can exceed acceptable levels when treating extensive tumours.
In such a case, the radiation exposure could be further decreased by irradiating from a different angle. Furthermore, by increasing the possible beam angle, more tumour sites could be treated. One option could be to use a design like the \emph{Planar Iso-centric System} (see Section~\ref{section:GantryLess}), with three beamline heads on the delivery system, each at a different fixed angle, allowing treatment from various angles. Such a setup is more expensive than a single-head setup, as each head would require a separate set of scanner magnets. 
However, it would offer more possibilities in treatment sites and radiation distribution. Such a setup is also supported by the findings of Susu Yan et al. \cite{yan2016reassessment}, who reported that, in their study, more than 50\% of the fields used to treat targets in the torso employed angles such as 0\degree, 90\degree, 180\degree, and 270\degree. Another option would be to have the patient sitting or in an upright position on a rotatable stool or platform, with the beam entering at a horizontal angle.
Custom restraints could support the patient to ensure repeated accurate positioning. As the treatment can be completed relatively quickly, the patient would not need to remain in such positions for an extended period. Such a setup would also increase the number of treatable sites without increasing the complexity and cost of the beam delivery system.

In Figures~\ref{fig:RadMap_SB}, \ref{fig:RadMap_SVB}, and \ref{fig:RadMap_ScannersXY}, we can also see the possible outcomes of the delivery plan under irregular breathing. It is interesting to see that the plan for the \emph{Scanner XY} setup seems to be the most susceptible to irregularities in the breathing pattern. The homogeneity of the dose distribution within the target volume appears worse for smaller tumours. 
This can mean that, at larger tumour sizes, the irregularities in breathing have more time to average out. This could also be why the scanner setups \emph{SB} and \emph{SVB} are less affected than the fast-paced \emph{Scanner XY} setup. Additionally, there is an increase in radiation in healthy tissues within the mobile part of the phantom. The stationary upper region of the phantom is not affected by the irregular breathing, and thus the distribution remains the same.

We can see that the tool is effective at creating a 4D plan for moving tumours, including position, delivery time, and optimised weight. The drawback, though, is its susceptibility to irregularities or non-predicted patterns of breathing. It would be possible to utilise ultrasound to observe the patient's breathing and synchronise with the planned procedure. 
In case of irregularities, the treatment could be paused and resumed when synchronised again. Another downside of the current version of 4D pencil beam delivery optimisation is the time-consuming weight optimisation.
With the increasing number of pencil beams to consider, optimisation can take hours. Furthermore, the grid positioning of the pencil beam might not be the best distribution of beams. Here, it would be interesting to explore the use of machine learning to reduce the time required for weight optimisation and investigate other beam positioning patterns for more effective delivery.

\section{Conclusion}
In this paper, we demonstrate that simulated breathing motion can be integrated into the treatment planning process. We tested different magnet-free and gantry-less setups with a simulated liver tumour affected by breathing motion. These simulations offer fascinating insights into magnet-free delivery systems. Removing both scanner magnets incurs a significant time cost during delivery. The cost reduction from removing the scanner magnets is less impactful than removing the gantry, and, in our opinion, insufficient to offset the increased delivery time, particularly since using a scanner magnet along the breathing axis reduces the time by at least 86\%. 
We found that the limiting factor is the accumulation of dose in healthy tissue. We conclude that building a system with some form of rotational \emph{DOF} could improve distribution while being more versatile and accommodating for different treatment sites. Such a system could comprise a multithreaded delivery system or a rotatable patient support system.
For the 4D planning tool, we propose incorporating machine learning into future work to speed up weight optimisation and explore more effective beam positioning patterns. 
Furthermore, given the treatment plan's sensitivity to irregular breathing patterns, we propose investigating the feasibility of a synchronisation tool in combination with an ultrasound device. Such a tool could allow the plan to synchronise with the current breathing pattern and, in severe cases of irregularity, pause the treatment and resume it once breathing returns to normal.

\newpage

\bibliographystyle{agsm}
\bibliography{magnetlesspt}

\end{document}


\title{Supplementary Material}

\date{}

\appendix
\noindent Supplementary Material referenced in the main article "Magnet-Free Proton Therapy with 4D Pencil Beam Delivery Optimisation".

\section{Optimised Beam Trajectory}\label{appendix:OptimisedBeamTrajecory}
\begin{figure} [H]
    \centering
    \begin{subfigure}{0.4\textwidth}
        \centering
        \includegraphics[width=\textwidth]{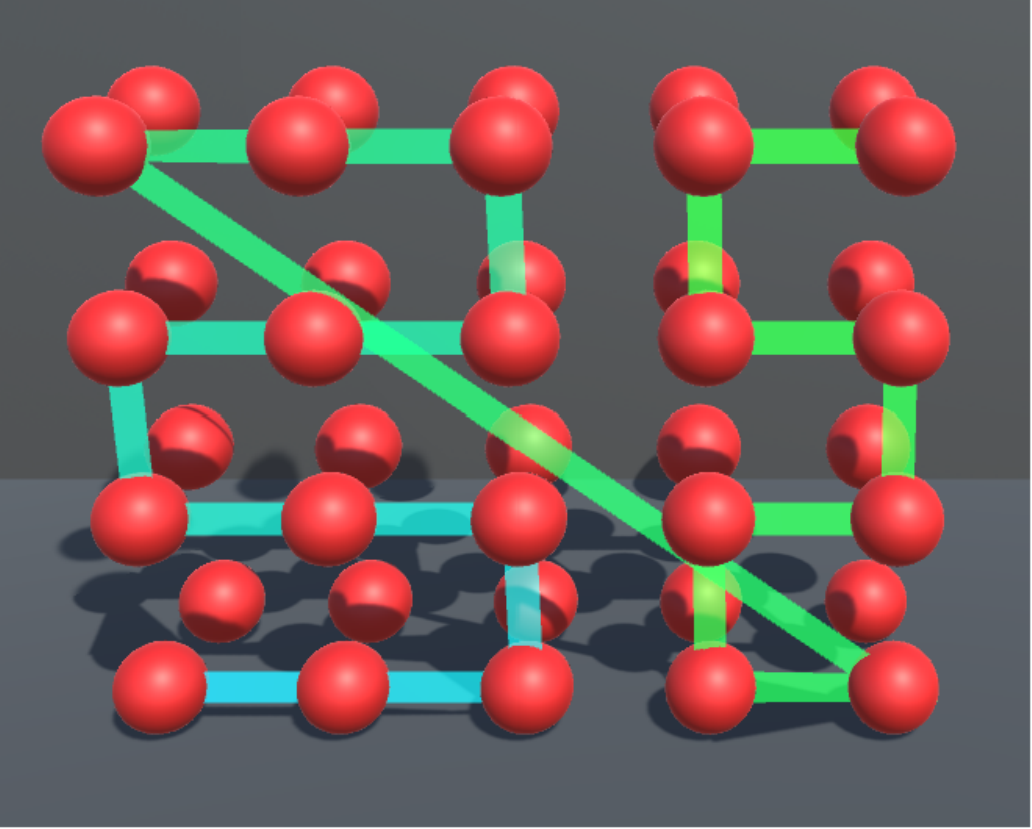}
        \caption{\emph{Stationary Beam}}\label{fig:nomove}
    \end{subfigure}%
    \hfill 
    \begin{subfigure}{0.4\textwidth}
        \centering
        \includegraphics[width=\textwidth]{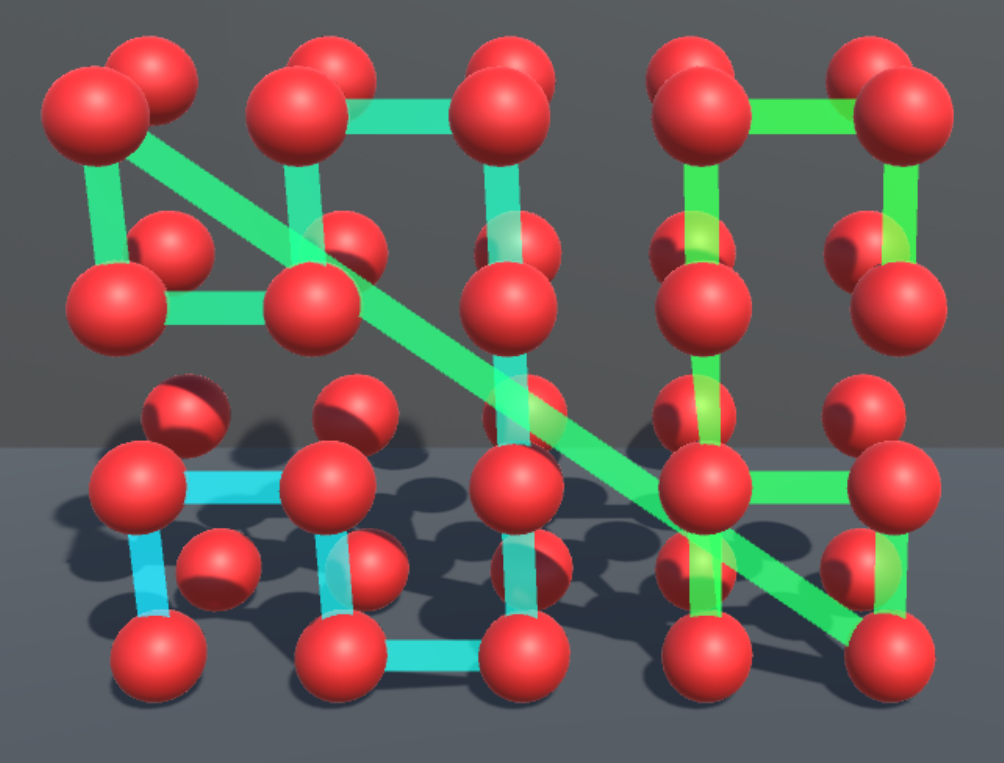}
        \caption{\emph{Stationary Variational Beam}}\label{fig:nomove_witheng}
    \end{subfigure}
    \vspace{1em} 
    \begin{subfigure}{0.4\textwidth}
        \centering
        \includegraphics[width=\textwidth]{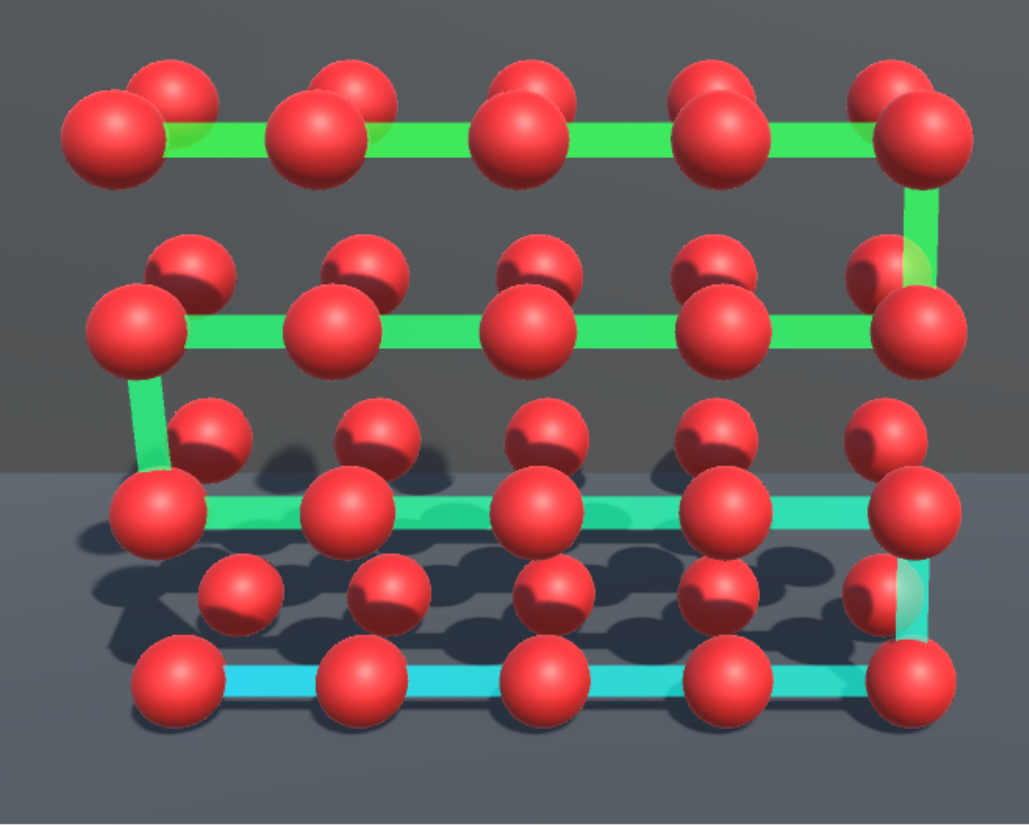}
        \caption{\emph{Scanner X}}\label{fig:x_scann}
    \end{subfigure}%
    \hfill 
    \begin{subfigure}{0.4\textwidth}
        \centering
        \includegraphics[width=\textwidth]{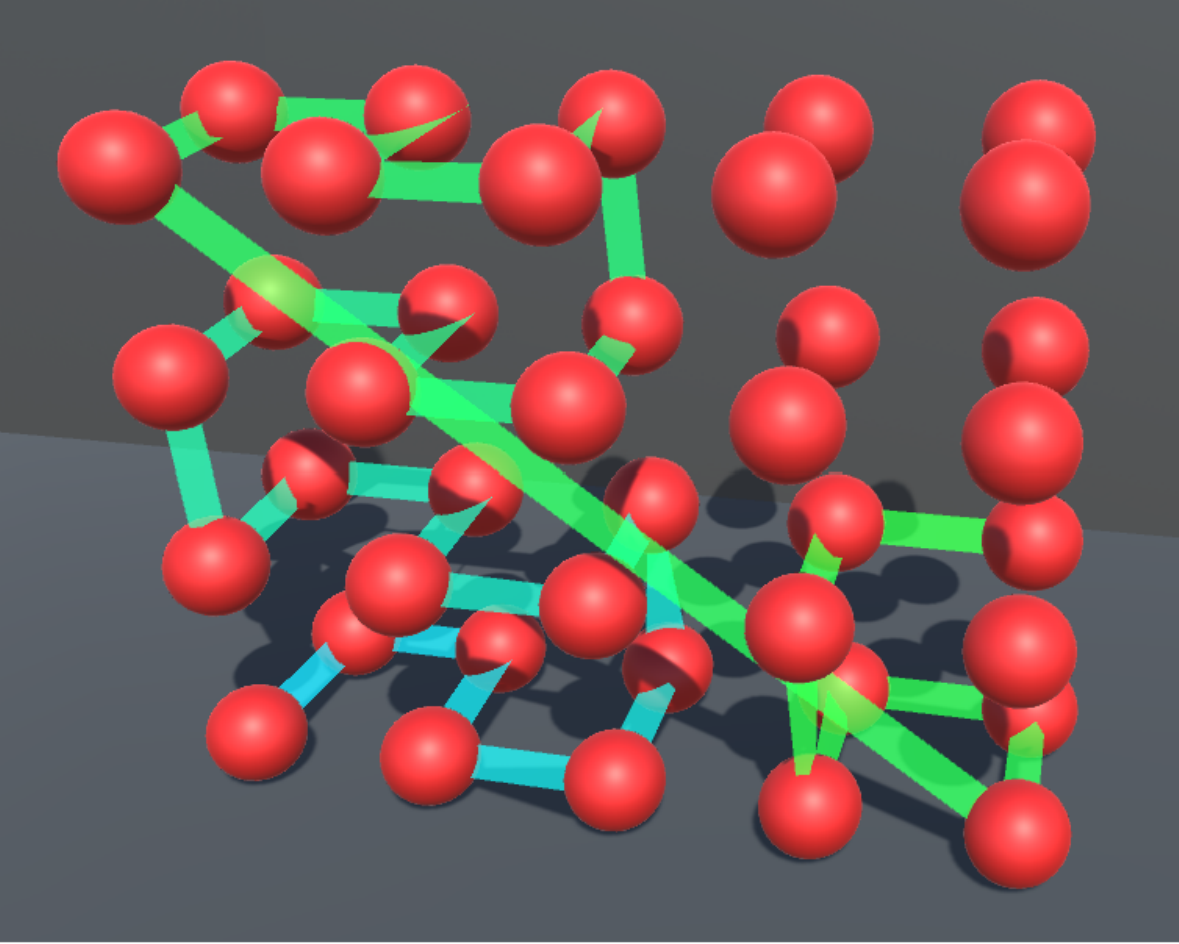}
        \caption{\emph{Scanner Y}}\label{fig:y_scann}
    \end{subfigure}
    \vspace{1em} 
    \begin{subfigure}{0.4\textwidth}
        \includegraphics[width=\textwidth]{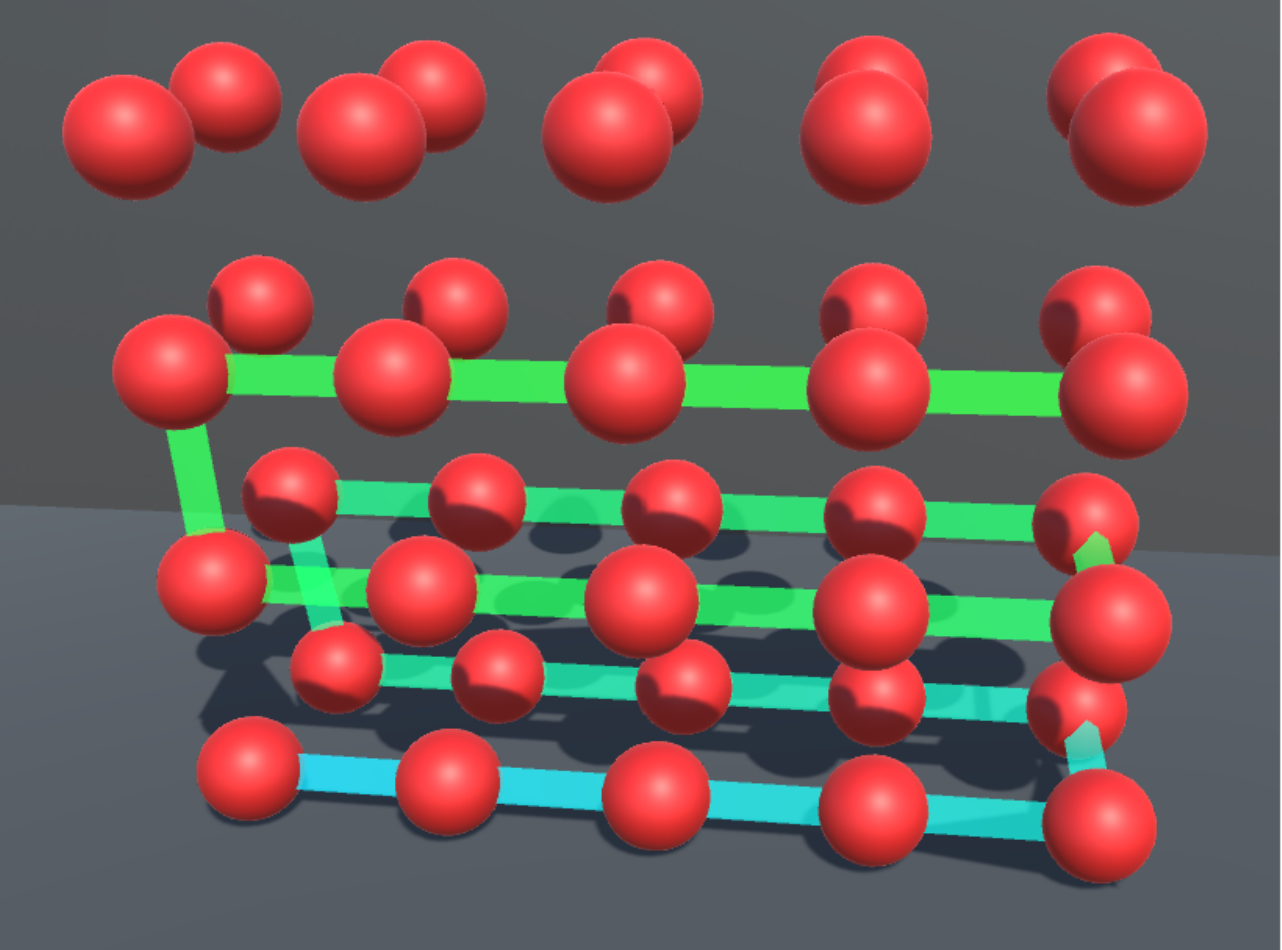}
        \caption{\emph{Scanner XY}}\label{fig:xy_scann}
    \end{subfigure}
    \hfill 
    \begin{subfigure}{0.4\textwidth}
        \centering
        \includegraphics[width=0.8\textwidth]{figures/figure1a.pdf}
        \caption{Schematic of the phantom.}\label{fig:phantoming} 
    \end{subfigure}
    \caption{ Illustration of the optimal path of dose deposition in gantry-free setups for the fastest treatment according to scanner properties and configurations in an exemplary grid of size $5\times2\times4$. The red sphere represents the centre of a spot, the blue-green line represents the order in which the positions are attended, and the blue part represents the path's beginning. (a) The example path for the \emph{Stationary Beam} PT Setup, without any scanner magnets, utilises only breathing and table motion. (b) Example pathway for \emph{Stationary Variational Beam} PT setup, without scanner magnets, but allowing for energy changes during the breathing cycle. (c) Example path for \emph{Scanner X} scanner setup, with one scanner magnet along the breathing axis. (d) Example path of the \emph{Scanner Y} treatment setup with one scanner magnet perpendicular to the breathing axis. (e) Example path of the \emph{Scanner XY} treatment setup with scanner magnets in both axes.} \label{fig:paths}
\end{figure}

Figure~\ref{fig:paths} shows schematic trajectories for the different beam delivery setups. The red sphere represents the centre of a spot. The blue-green line represents the order in which the positions are attended, with the blue part representing the beginning of the path. Figure~\ref{fig:phantoming} shows the respective coordinate system within the phantom. The horizontal axis is the $x$-axis or the breathing axis, the vertical axis is the $z$-axis along the beam, and the $y$-axis is the depth.

For the \emph{SB} and \emph{SVB} setups, we removed the gantry and scanner magnets. We rely only on the patient's breathing and table movement to scan the tumour. The proposed two scanning paths are illustrated in Figure~\ref{fig:nomove} and Figure~\ref{fig:nomove_witheng} for the \emph{SB} and \emph{SVB} setups, respectively. In the \emph{SB} setup, Figure~\ref{fig:nomove}, we propose that the target is moved through the beam by breathing, essentially self-scanning, and the beam energy change occurs after inhaling or exhaling. The table advances to the next slice after an entire section has been treated. For the \emph{SVB} setup, Figure~\ref{fig:nomove_witheng}, the target moves through the beam by breathing, and the energy changes occur during in- or exhalation, essentially allowing the treatment of four rows in one breathing cycle. 

The \emph{Scanner X} and \emph{Scanner Y} setups each have one scanner magnet, either along the $x$-axis (meaning along the breathing axis), as shown in Figure~\ref{fig:x_scann} or one along the $y$-axis (perpendicular to the breathing axis), as shown in Figure~\ref{fig:y_scann}. For the \emph{Scanner X} setup, the scanner magnet can scan the tumour in both directions. Thus, the motion is not needed to scan the tumour, but must be accounted for in the scanner magnet's movement. The table moves the patient along the $y$-axis to the next section within the target. The \emph{Scanner Y} setup utilises the brething motion to scan in the $x$-axis while the scanner magnet moves the beam along the $y$-axis. If the breathing depth is insufficient to cover the entire target, the table moves the target along the $x$-axis to the next section.

The \emph{Scanner XY} setup in Figure~\ref{fig:xy_scann} has both scanner magnets. Neither the table nor the breathing motion is required to move the target through the beam. However, the breathing motion is accounted for in the scanner magnets' motion.

\section{Weight Optimisation}
\subsection{Lookup Table - WED to RSP}\label{appendix:RSP}
\begin{figure}[H]
    \centering
    \includegraphics[width=1\linewidth]{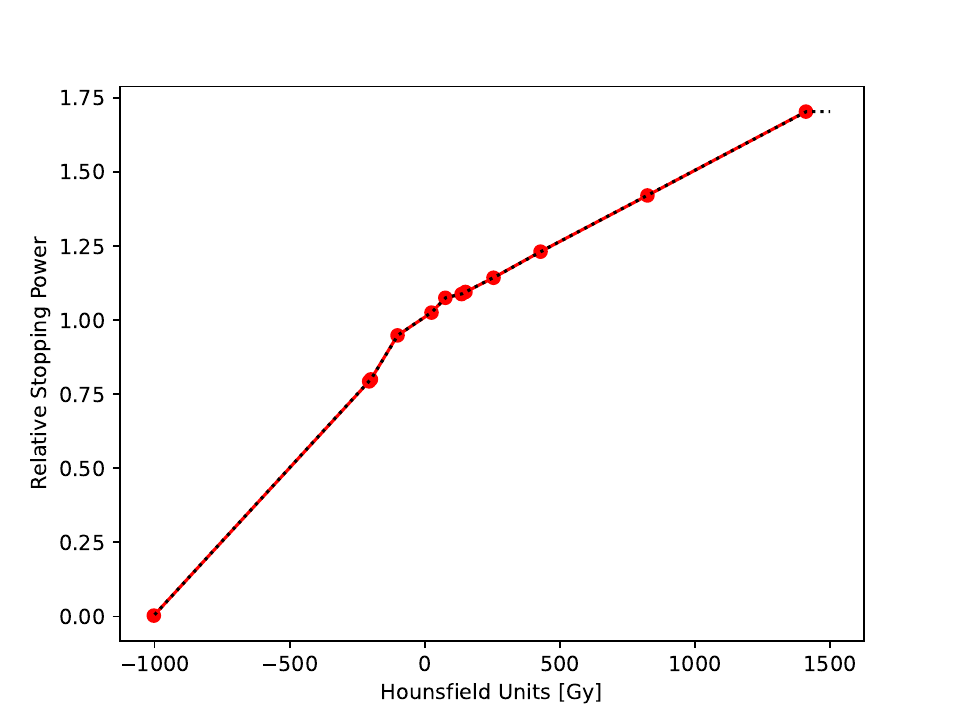}
    \caption{Lookup Table for transforming Hounsfield Units into relative stopping power.}\label{fig:HUtoRSP}
\end{figure}

Figure~\ref{fig:HUtoRSP} shows the correlation between HU to RSP. This table was used to transform the phantom's CT into RSP (van Abbema et al. 2018).

\subsection{Beam Calculation} \label{appendix:Beamcalc}
\begin{figure}[H]
    \centering
    \includegraphics[width=1\linewidth]{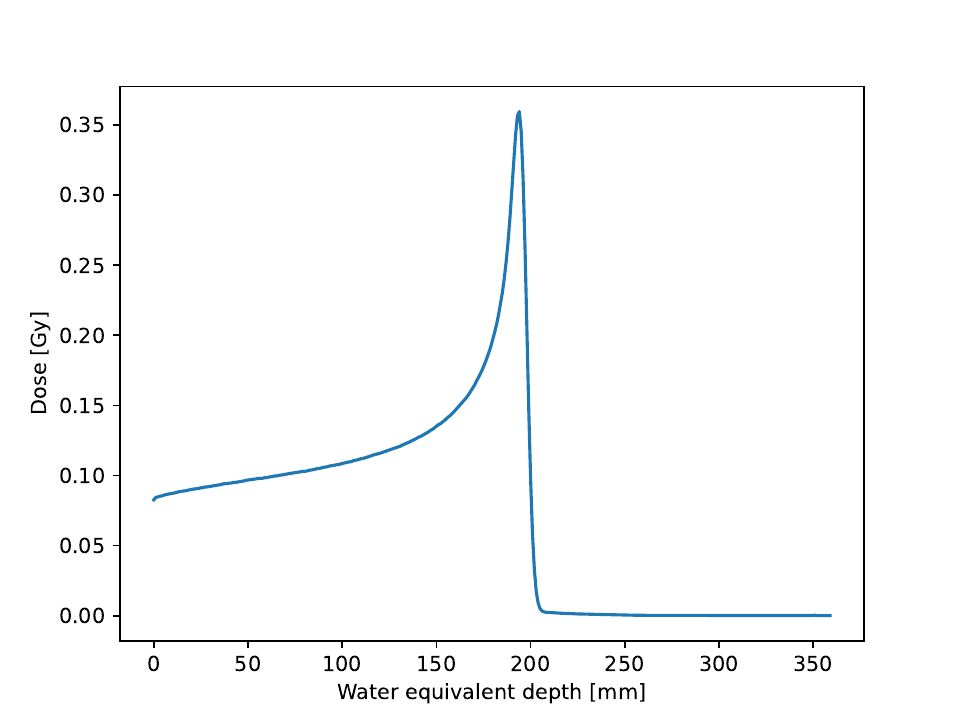}
    \caption{Integral dose, interpolated from the depth-dose-look-up table, given by the PSI.}\label{fig:ID}
\end{figure}

\begin{figure}[H]
    \centering
    \includegraphics[width=1\linewidth]{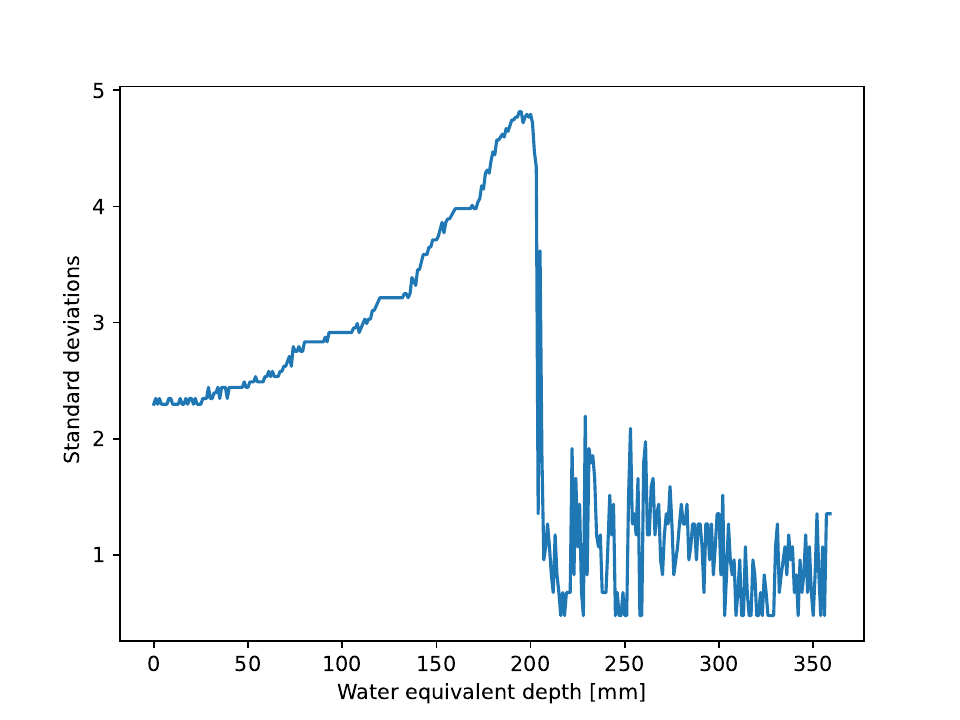}
    \caption{Standard deviation, given by the PSI.}\label{fig:SD}
\end{figure}

Figure~\ref{fig:ID} and \ref{fig:SD} show the integral dose and respective standard deviation obtained from the beam given by the PSI. They were utilised to calculate the beam in the water-equivalent range.

\subsection{Optimisation Metrics} \label{appendix:Optimisation}
Listed in Table~\ref{tab:TraningMetrics} are the metrics used for optimising the weights. The same metrics were used for all experiments.

\begin{table*}[h]
\caption{Parameters used in the optimisation of the weights.}
\label{tab:TraningMetrics}
\centering
\begin{tabular}{l|ccc}\hline
\toprule
\textbf{Metric} & $\mathbf{\SI{30}{mm}}$ & $\mathbf{\SI{40}{mm}}$ & $\mathbf{\SI{50}{mm}}$ \\ 
\midrule
Iterations & 40'000 & 40'000 & 40'000 \\
Learning Rate& 0.05 & 0.05 & 0.05 \\ 
Weight $c_{max}$ & 0.5 & 0.5 & 0.5 \\ 
\bottomrule
\end{tabular}
\end{table*}

\begin{figure}[H]
    \centering
    \includegraphics[width=0.8\linewidth]{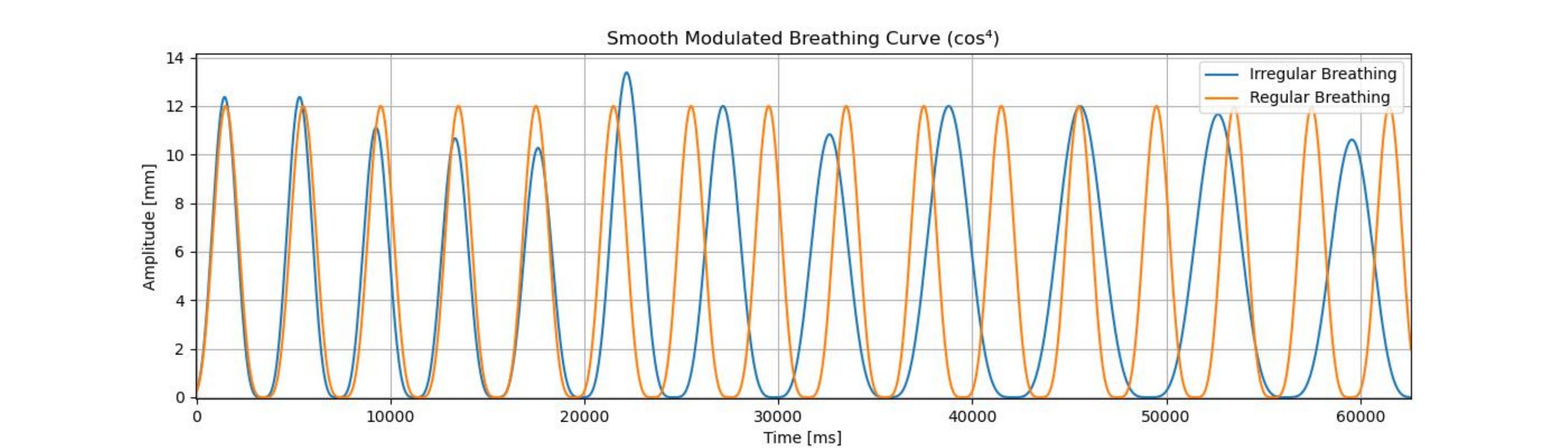}
    \caption{Example of an irregular breathing pattern: in orange the regular breathing and in blue the irregular breathing pattern (blue).}\label{fig:irregular_pattern_example}
\end{figure}

Figure~\ref{fig:irregular_pattern_example} shows an example of an irregular breathing pattern. Both amplitude and frequency are modulated randomly.

\section{Additional results}\label{appendix:Results}

\subsection{Delivery Time}\label{appendix:DeliveryTime}
\begin{figure}[H]
    \centering
    \includegraphics[width=1\linewidth]{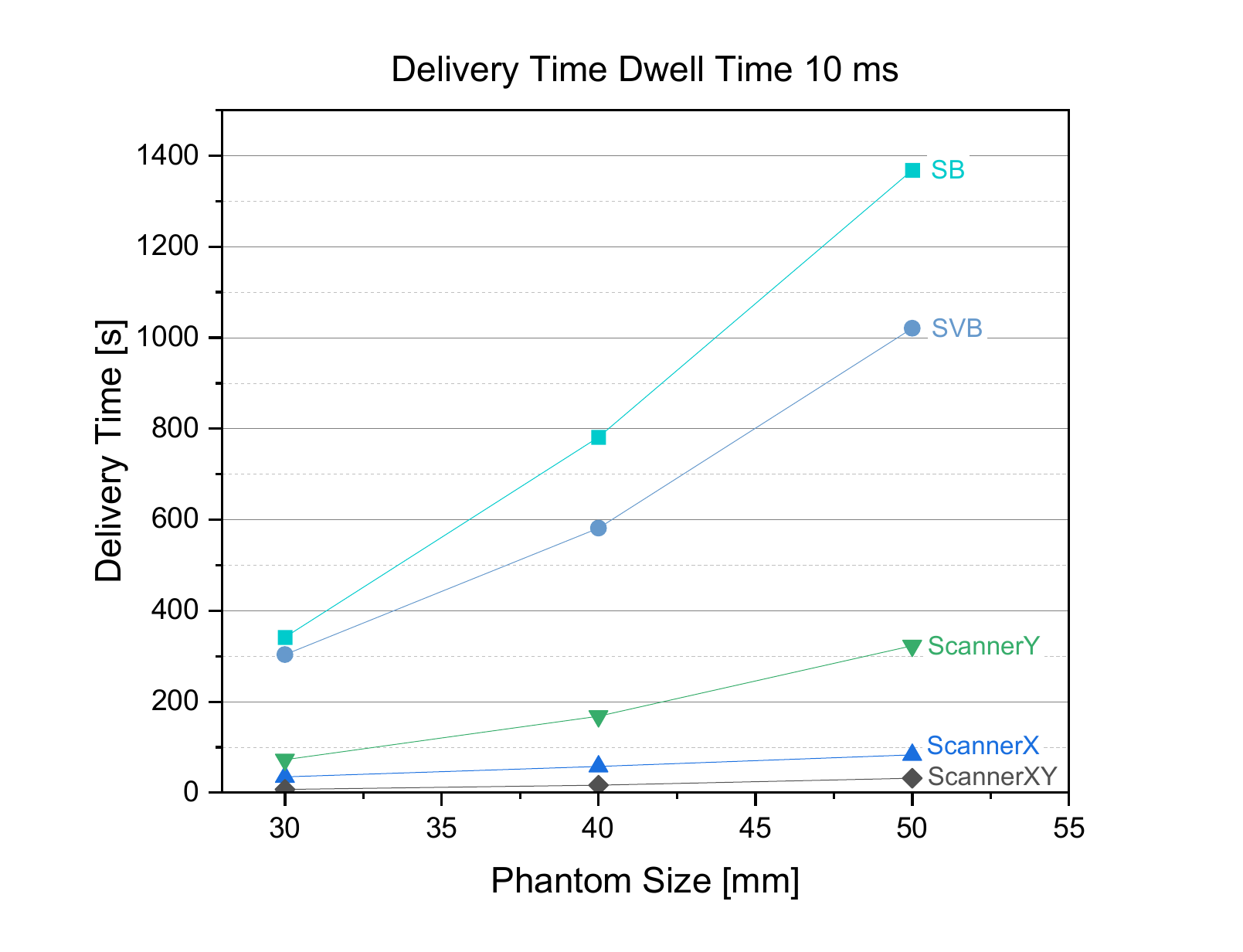}
    \caption{Delivery time for a dwell time of \SI{10}{ms} plotted against the target volume size.}\label{fig:deliverytime_toPS}
\end{figure}

Figure~\ref{fig:deliverytime_toPS} shows the delivery time for a dwell time of \SI{10}{ms} plotted against the target volume size.

\subsection{Metrics of optimised radiation Maps.}\label{appendix:ResultMetrics}
This section shows the metrics for all the experiments conducted. Tables~\ref{tab:CompleteMetrics_SB_regular} and \ref{tab:CompleteMetrics_SB_irregular} include all the metrics for the scanner variant \emph{SB} for regular breathing and irregular breathing, respectively. Tables~\ref{tab:CompleteMetrics_SVB_regular} and \ref{tab:CompleteMetrics_SVB_irregular} hold the metrics for the \emph{SVB} scanner mode, and tables~\ref{tab:CompleteMetrics_ScannersXY_regular} and \ref{tab:CompleteMetrics_ScannersXY_irregular} have the metrics for \emph{Scanners XY} delivery method.

\begin{table*}[htbp]
\caption{The calculated metrics for \SI{30}{mm}, \SI{40}{mm} and \SI{50}{mm} phantom spheres for \emph{SB} scanner mode under regular breating. There are also the respective optimal values (OV) used for the optimisation process.}
\label{tab:CompleteMetrics_SB_regular}
\centering
\begin{tabular}{l|cccc}
\toprule
\textbf{SB reg} & $\mathbf{\SI{30}{mm}}$ & $\mathbf{\SI{40}{mm}}$ & $\mathbf{\SI{50}{mm}}$ & \textbf{OV} \\
\midrule
$\mathrm{D_{98 \%}}$ [Gy] & 8.49 & 8.85 & 8.56 & \\
$\mathrm{D_{95 \%}}$ [Gy] & 9.16 & 9.51 & 9.28 & 10 \\
$\mathrm{D_{50 \%}}$ [Gy] & 10.51 & 10.46 & 10.29 & \\
$\mathrm{D_{2 \%}}$ [Gy] & 11.92 & 11.91 & 11.91 & \\
\midrule
$\mathrm{D_{max}HT}$ [Gy] & 9.64 & 9.89 & 10.02 & 7 \\ 
$\mathrm{D_{max}TV}$ [\%] & 119.99 & 119.98 & 120.00 & 120 \\ 
\midrule
$\mathrm{V_{100 \%}}$ [\%] & 64.72 & 67.14 & 60.65 & \\
$\mathrm{V_{98 \%}}$ [\%] & 71.00 & 79.14 & 69.25 & \\ 
$\mathrm{V_{95 \%}}$ [\%] & 82.60 & 95.04 & 84.72 & 98 \\ 
\midrule
CN & 0.81 & 0.87 & 0.77 & 1 \\ 
HI & 0.33 & 0.29 & 0.33 & 0\\
\bottomrule
\end{tabular}
\end{table*}

\begin{table*}[htbp]
\caption{The calculated metrics for \SI{30}{mm}, \SI{40}{mm} and \SI{50}{mm} phantom spheres for \emph{SB} scanner mode under irregular breating. There are also the respective optimal values (OV) used for the optimisation process.}
\label{tab:CompleteMetrics_SB_irregular}
\centering
\begin{tabular}{l|cccc}
\toprule
\textbf{SB irreg} & $\mathbf{\SI{30}{mm}}$ & $\mathbf{\SI{40}{mm}}$ & $\mathbf{\SI{50}{mm}}$ & \textbf{OV} \\
\midrule
$\mathrm{D_{98 \%}}$ [Gy] & 6.04 & 7.03 & 7.20 & \\
$\mathrm{D_{95 \%}}$ [Gy] & 6.63 & 7.65 & 7.71 & 10 \\
$\mathrm{D_{50 \%}}$ & 9.17 & 9.58 & 9.51 & \\
$\mathrm{D_{2 \%}}$ [Gy] & 12.04 & 11.43 & 12.38 & \\
\midrule
$\mathrm{D_{max}HT}$  [Gy] & 10.68 & 10.27 & 11.37 & 7 \\ 
$\mathrm{D_{max}TV}$ [\%] & 131.63 & 125.22 & 136.91 & 120 \\ 
\midrule
$\mathrm{V_{100 \%}}$ [\%] & 26.59 & 32.63 & 35.08 & \\
$\mathrm{V_{98 \%}}$[\%] & 31.89 & 41.82 & 40.58 & \\ 
$\mathrm{V_{95 \%}}$ [\%] & 40.28 & 52.87 & 50.39 & 98 \\ 
\midrule
CN & 0.37 & 0.50 & 0.45 & 1 \\ 
HI & 0.65 & 0.46 & 0.55 & 0\\
\bottomrule
\end{tabular}
\end{table*}

\begin{table*}[htbp]
\caption{The calculated metrics for \SI{30}{mm}, \SI{40}{mm} and \SI{50}{mm} phantom spheres for \emph{SVB} scanner mode under regular breating. There are also the respective optimal values (OV) used for the optimisation process.}
\label{tab:CompleteMetrics_SVB_regular}
\centering
\begin{tabular}{l|cccc}
\toprule
\textbf{SVB reg} & $\mathbf{\SI{30}{mm}}$ & $\mathbf{\SI{40}{mm}}$ & $\mathbf{\SI{50}{mm}}$ & \textbf{OV} \\
\midrule
$\mathrm{D_{98 \%}}$ [Gy] & 8.62 & 8.85 & 8.90 & \\
$\mathrm{D_{95 \%}}$ [Gy] & 9.22 & 9.51 & 9.57 & 10 \\
$\mathrm{D_{50 \%}}$ [Gy] & 10.56 & 10.38 & 10.36 & \\
$\mathrm{D_{2 \%}}$ [Gy] & 11.94 & 11.91 & 11.92 & \\
\midrule
$\mathrm{D_{max}HT}$ [Gy] & 9.73 & 9.92 & 9.94 & 7 \\ 
$\mathrm{D_{max}TV}$ [\%] & 120.0 & 119.97 & 119.99 & 120 \\ 
\midrule
$\mathrm{V_{100 \%}}$ [\%] & 66.33 & 64.44 & 66.26 & \\
$\mathrm{V_{98 \%}}$ [\%] & 73.14 & 77.17 & 78.15 & \\ 
$\mathrm{V_{95 \%}}$ [\%] & 84.99 & 95.08 & 95.36 & 98 \\ 
\midrule
CN & 0.82 & 0.86 & 0.87 & 1 \\ 
HI & 0.32 & 0.30 & 0.29 & 0\\
\bottomrule
\end{tabular}
\end{table*}

\begin{table*}[htbp]
\caption{The calculated metrics for \SI{30}{mm}, \SI{40}{mm} and \SI{50}{mm} phantom spheres for \emph{SVB} scanner mode under irregular breating. There are also the respective optimal values (OV) used for the optimisation process.}
\label{tab:CompleteMetrics_SVB_irregular}
\centering
\begin{tabular}{l|cccc}
\toprule
\textbf{SVB irreg} & $\mathbf{\SI{30}{mm}}$ & $\mathbf{\SI{40}{mm}}$ & $\mathbf{\SI{50}{mm}}$ & \textbf{OV} \\
\midrule
$\mathrm{D_{98 \%}}$ [Gy] & 6.41 & 7.01 & 6.82 & \\
$\mathrm{D_{95 \%}}$ [Gy] & 7.09 & 7.49 & 7.46 & 10 \\
$\mathrm{D_{50 \%}}$ [Gy] & 9.14 & 9.86 & 9.37 & \\
$\mathrm{D_{2 \%}}$ [Gy] & 11.29 & 13.51 & 12.65 & \\
\midrule
$\mathrm{D_{max}HT}$ [Gy] & 11.01 & 11.51 & 11.89 & 7 \\ 
$\mathrm{D_{max}TV}$ [\%] & 118.01 & 146.39 & 151.12 & 120 \\ 
\midrule
$\mathrm{V_{100 \%}}$ [\%] & 23.19 & 49.65 & 30.25 & \\
$\mathrm{V_{98 \%}}$ [\%] & 28.73 & 51.48 & 35.66 & \\ 
$\mathrm{V_{95 \%}}$ [\%] & 37.73 & 58.78 & 45.48 & 98 \\ 
\midrule
CN & 0.36 & 0.53 & 0.40 & 1 \\ 
HI & 0.53 & 0.66 & 0.62 & 0\\
\bottomrule
\end{tabular}
\end{table*}

\begin{table*}[htbp]
\caption{The calculated metrics for \SI{30}{mm}, \SI{40}{mm} and \SI{50}{mm} phantom spheres for \emph{Scanners XY} scanner mode under regular breathing. There are also the respective optimal values (OV) used for the optimisation process.}
\label{tab:CompleteMetrics_ScannersXY_regular}
\centering
\begin{tabular}{l|cccc}
\toprule
\textbf{S. XY reg} & $\mathbf{\SI{30}{mm}}$ & $\mathbf{\SI{40}{mm}}$ & $\mathbf{\SI{50}{mm}}$ & \textbf{OV} \\
\midrule
$\mathrm{D_{98 \%}}$ [Gy] & 8.32 & 8.88 & 8.85 & \\
$\mathrm{D_{95 \%}}$ [Gy] & 9.14 & 9.56 & 9.52 & 10 \\
$\mathrm{D_{50 \%}}$ [Gy] & 10.6 & 10.41 & 10.37 & \\
$\mathrm{D_{2 \%}}$ [Gy] & 11.93 & 11.92 & 11.90 & \\
\midrule
$\mathrm{D_{max}HT}$ [Gy] & 9.72 & 9.97 & 9.95 & 7 \\ 
$\mathrm{D_{max}TV}$ [\%] & 120.00 & 120.02 & 120.0 & 120 \\ 
\midrule
$\mathrm{V_{100 \%}}$ [\%] & 70.21 & 68.99 & 67.17 & \\
$\mathrm{V_{98 \%}}$ [\%] & 76.85 & 81.71 & 78.98 & \\ 
$\mathrm{V_{95 \%}}$ [\%] & 87.53 & 95.44 & 95.15 & 98 \\ 
\midrule
CN & 0.85 & 0.85 & 0.86 & 1 \\ 
HI & 0.34 & 0.29 & 0.30 & 0\\
\bottomrule
\end{tabular}
\end{table*}

\begin{table*}[htbp]
\caption{The calculated metrics for \SI{30}{mm}, \SI{40}{mm} and \SI{50}{mm} phantom spheres for \emph{Scanners XY} scanner mode under irregular breating. There are also the respective optimal values (OV) used for the optimisation process.}
\label{tab:CompleteMetrics_ScannersXY_irregular}
\centering
\begin{tabular}{l|cccc}
\toprule
\textbf{S. XY irreg} & $\mathbf{\SI{30}{mm}}$ & $\mathbf{\SI{40}{mm}}$ & $\mathbf{\SI{50}{mm}}$ & \textbf{OV} \\
\midrule
$\mathrm{D_{98 \%}}$ [Gy] & 5.57 & 5.52 & 6.18 & \\
$\mathrm{D_{95 \%}}$ [Gy] & 6.30 & 6.08 & 6.73 & 10 \\
$\mathrm{D_{50 \%}}$ [Gy] & 9.21 & 9.61 & 9.44 & \\
$\mathrm{D_{2 \%}}$ [Gy] & 11.66 & 13.42 & 12.76 & \\
\midrule
$\mathrm{D_{max}HT}$ [Gy] & 10.20 & 11.53 & 12.91 & 7 \\ 
$\mathrm{D_{max}TV}$ [\%] & 122.06 & 151.13 & 149.98 & 120 \\ 
\midrule
$\mathrm{V_{100 \%}}$ [\%] & 27.10 & 42.17 & 34.39 & \\
$\mathrm{V_{98 \%}}$ [\%] & 33.53 & 45.96 & 39.83 & \\ 
$\mathrm{V_{95 \%}}$ [\%] & 42.16 & 52.13 & 48.22 & 98 \\ 
\midrule
CN & 0.40 & 0.43 & 0.40 & 1 \\ 
HI & 0.66 & 0.82 & 0.70 & 0\\
\bottomrule
\end{tabular}
\end{table*}